\documentclass[journal]{IEEEtran}

\IEEEoverridecommandlockouts                              

\usepackage{graphics} 
\usepackage{epsfig} 

\usepackage{amsmath} 
\usepackage{amssymb}  
\usepackage{mathtools} 
\usepackage{bm} 
 \usepackage{multirow}
 \usepackage[linesnumbered,ruled,vlined,Algorithm ]{algorithm }
\usepackage{algpseudocode}
\usepackage{hyperref}
\usepackage{array}

\usepackage{epstopdf}

\usepackage{subcaption}
\usepackage{enumerate}

\usepackage[english]{babel} 
\usepackage{amsfonts,amsthm,bm} 
 \usepackage{cleveref}

\usepackage[latin1]{inputenc}
\usepackage{tikz}
\usetikzlibrary{shapes,arrows}

\usepackage{graphicx}
\usepackage{wrapfig }
\usepackage{multirow}
 \usepackage[edges]{forest}

\usepackage{tcolorbox}
\usepackage{tabularx,colortbl}
\usepackage{adjustbox}
\usepackage{colortbl}
\usepackage{hhline}

\DeclareMathOperator*{\argmin}{arg\,min}

 \usepackage{cite}
  
  \usepackage{stfloats}

\title{\LARGE \bf
Priority-based DREAM Approach for Highly Manoeuvring Intruders in A Perimeter Defense Problem
}

\author{Shridhar Velhal$^{1}$, Suresh Sundaram$^{2}$ and Narasimhan Sundararajan$^{3}$ 
\thanks{$^{1}$Shridhar Velhal is a PhD student at Department of Aerospace Engineering, Indian Institute of Science, Bengaluru, India.       {\tt\small velhalb@iisc.ac.in}}
\thanks{$^{2}$Suresh Sundaram is an Associate Professor at Department of Aerospace Engineering, Indian Institute of Science, Bengaluru, India.{\tt\small vssuresh@iisc.ac.in}}%

\thanks{$^{3}$N. Sundararajan is research staff in WIRIN project, Department of Aerospace Engineering, Indian Institute of Science, Bengaluru, India.   {\tt\small ensundara@gmail.com}}%
}

\begin{document}

\maketitle

\begin{abstract}
In this paper, a Priority-based Dynamic REsource Allocation with decentralized Multi-task assignment (P-DREAM) approach is presented to protect a territory from highly manoeuvring intruders. In the first part, static optimization problems are formulated to compute the following parameters of the perimeter defense problem; the number of reserve stations, their locations, the priority region, the monitoring region, and the minimum number of defenders required for the monitoring purpose. The concept of a prioritized intruder is proposed here to identify and handle those critical intruders (computed based on the velocity ratio and location) to be tackled on a priority basis. The computed priority region helps to assign reserve defenders sufficiently earlier such that they can neutralize the prioritized intruders. The monitoring region defines the minimum region to be monitored and is sufficient enough to handle the intruders. In the second part, the earlier developed DREAM approach is modified to incorporate the priority of an intruder. The proposed P-DREAM approach assigns the defenders to the prioritized intruders as the first task. A convex territory protection problem is simulated to illustrate the P-DREAM approach. It involves the computation of static parameters and solving the prioritized task assignments with dynamic resource allocation. Monte-Carlo results were conducted to verify the performance of P-DREAM, and the results clearly show that the P-DREAM approach can protect the territory with consistent performance against highly manoeuvring intruders.
\end{abstract}


\section{Introduction}
 
\IEEEPARstart{W}{ith} the recent advances in technologies, Unmanned Aerial Vehicles (UAVs) have found diverse applications in areas such as agriculture, logistics, healthcare, urban mobility, and defense \cite{shakhatreh2019unmanned}. The volume of UAV traffic is predicted to rise enormously soon. With the autonomy, UAVs also threaten the privacy, safety, and security of safety-critical infrastructures \cite{gusterson2016drone}. This demands airspace protection systems for protecting lower airspace from UAVs \cite{kang2020protect}. In practice, one has authority only for a critical territory and the surrounding region may be unrestricted. A team of defenders needs to protect the territory (while operating inside and on the territory) by neutralizing the intruders (on the perimeter) before an intruder enters the territory; this problem is referred to as a Perimeter Defense Problem (PDP). A detailed review of PDP is given  in \cite{shishika2020review}. 

PDP requires a defense strategy for a team of defenders to neutralize the intruders (while operating on and inside the perimeter) before the intruders cross the perimeter. PDP can be cast as one form of a  differential game introduced by \cite{isaacs1999differential}.  PDP is a special case of  multiplayer reach and avoid games \cite{garcia2020multiple}, in which defenders are forced to operate inside the territory.  PDP was first proposed in \cite{agmon2008multi} as a perimeter patrolling problem. PDP has been solved for a linear boundary \cite{smith2009dynamic}, a  circular territory \cite{ bajaj2019dynamic}, a convex territory \cite{shishika2018local}, and  a 3-D spherical territory \cite{lee2020perimeter}.

In \cite{velhal2021decentralized},  one-to-many assignment scheme was used to assign multiple slowly manoeuvring intruders to each defender. \cite{velhal2021decentralized} used a predictive approach to predict the arrival locations of the intruders and then solved a spatio-temporal multi-task assignment problem.  A one-to-many assignment scheme was used to assign multiple slowly manoeuvring intruders to each defender. 
Further, the above study was extended to varying numbers of defenders, and Dynamic REsource Allocation with Multi-task assignments (DREAM) algorithm has been developed in \cite{velhal2022dynamic}
Here reserve stations were proposed inside the territory to facilitate a dynamic team size for the defenders.   \cite{chopra2014heterogeneous, chopra2015spatio} have computed the minimum number of defenders required to execute the given spatio-temporal tasks but the approach requires one  to solve the task assignment problem (Hungarian algorithm) multiple times in an iterative fashion.
However, DREAM approach \cite{velhal2022dynamic} provided a one-step solution (task assignment problem is solved only once) to compute the minimum number of required defenders. 
After that, the task assignment solution was recomputed for the updated optimal team of defenders to guarantee the execution of all the spatio-temporal tasks by utilizing the minimum number of defenders.
Recently, \cite{adler2022role} proposed a dynamic programming approach to compute the assignments of a spatio-temporal task assignment problem defined by PDP for a circular territory with two heterogeneous defenders. The above problem formulations based on the estimation of arrival location was valid only for the case of slowly manoeuvring intruders. In earlier studies, it was found that the success rate of the DREAM approach reduced significantly for highly manoeuvring intruders. Hence there is a strong need to develop a strategy for defending a territory against highly manoeuvring intruders with optimal utilization of resources and this paper attempt to fill this gap.

This paper presents a priority-based dynamic resource allocation scheme with multi-task assignments (P-DREAM) approach for neutralizing highly manoeuvring multiple intruders in PDP. In the proposed approach, a team of defenders will continuously monitor the region of interest to detect and track the intruders. To improve the success in the neutralization of the intruders, those intruders who are sufficiently closer to the territory are handled on priority, as the first task. A defender will act according to the actions of the first assigned intruder and neutralize it. Meanwhile, subsequently assigned intruders may become infeasible, but these subsequent intruders are sufficiently away from the territory and can be reassigned. The priority of intruders is computed based on the location of that intruder, speed ratio, location of reserve stations, and also the shape of the territory. One needs to continuously monitor a monitoring region (inclusive of a priority region) to detect any potential intruder earlier. The monitoring region is computed based on the shape of the territory, worst-case speed ratio, and the monitoring sensor range.  After computing the monitoring region, the proposed approach identifies a minimum number of defenders required for the monitoring tasks based on the sensor ranges. The number of reserve stations is then optimized to reduce the size of this monitoring region.  All the above optimization problems are static in nature and need to be computed only once for a given convex territory. The computed number of reserve stations, their optimal locations, priority region, monitoring region, and the minimum number of defenders for monitoring are used in online assignment computation in PDP.

 Once the above static optimizations are solved and the PDP parameters are finalized, next, the priority-based spatio-temporal multi-task assignment method for assigning defenders to intruders while considering the priority of intruders is done. The cost function is modified to enforce the prioritized intruders as the first task.  Furthermore dynamic resource allocation algorithm is modified to incorporate the prioritization of the intruders, and a continuously monitoring task (with a specified monitoring region), and this approach is referred to as P-DREAM.  P-DREAM formulates an optimization problem that needs to be solved online, and has the same computational complexity as that of the DREAM approach.
 
Detailed illustration of the P-DREAM approach is given in the results section. First, the static PDP parameters are computed for a given territory.  As a first step, the number of reserve stations is selected, and based on this the priority and monitoring regions are computed. Next, based on the selected sensor ranges, the minimum number of defenders required for monitoring is computed. Based on the computed priority and monitoring region, the priority of intruders is decided. P-DREAM computes the cost matrix considering the priority and feasibility of spatio-temporal tasks. The P-DREAM approach first computes the minimum resources required for a given spatio-temporal tasks and deploys minimum defenders to execute those tasks. The spatio-temporal task assignment algorithm is again solved with an updated team of defenders to ensure that all prioritized tasks are assigned as first tasks and all assignments are feasible. The performance of this setup is evaluated using Monte-Carlo simulations. The P-DREAM approach shows consistent performance above $95\%$ with 3 reserve stations and unit speed ratio  against six active intruders with manoeuvrability up to $45 deg/sec$, whereas for the same scenario the DREAM approach fails to protect territory.

The rest of the paper is organized as follows: Section \ref{sec:PDP} presents the mathematical problem formulation of a typical perimeter defense problem. Section \ref{sec:static} provides the static problem formulations for determining optimal locations of reserve stations, computation of the priority and monitoring regions, and  the minimum number of defenders required for monitoring. Section \ref{sec:dynamic} describes the P-DREAM approach with priority region and monitoring task. Section \ref{sec:result} provides the performance evaluation results for P-DREAM using simulation studies .  Finally, Section \ref{sec:conclusion}  summarises the conclusions from this study.

\section{Related Works} \label{sec:lit_review}
In \cite{guerrero2020perimeter}  surveillance of  a perimeter has been guaranteed with an analytical solution using zeroing barrier functions for the case of defending the territory against a single intruder. This has been further extended to guarantee the protection of a non-convex region against a fast-moving intruder using multiple defenders\cite{guerrero2021robust}. Both these works were restricted to the case of a single intruder and multiple defenders. The sensor placement problem for surveillance purpose has been studied in \cite{feng2021sensor}. A 
The analytical barrier function-based solutions using territory geometry and the initial positions of defenders were presented in ~\cite{yan2019reach,yan2020task,yan2020guarding} for two defenders and one intruder games. Defender-attacker game for defending a single object has been studied by considering time and resources in ~\cite{zhang2019defending}.

Most of the works on multiplayer perimeter defense problems for highly maneuvering intruders mainly focus on a one-to-one assignment of a defender to an intruder \cite{shishika2018local,shishika2020cooperative}. The solutions are based on geometric methods where  the feasibility regions are computed first  and then  the assignment problem for feasible intruders is solved. If an intruder lies in an infeasible region, then that intruder can enter the territory irrespective of the defenders' actions.  These feasible/infeasible regions are based on the positions of all defenders in a team and vary with the positions of the defenders.  In \cite{shishika2019team}, a team of patrollers was separately used to monitor the intruders outside the territory; but the theoretically computed minimum number of patrollers and defenders was on the conservative side. Thus, the above works on PDP are of limited use  in practice and they have to be extended to handle constraints in the areas of defender dynamics, sequential capture of intruders, fast manoeuvring intruders, and availability of partial information \cite{shishika2020review}.

Recently, to protect a circular territory from radially incoming intruders with constant velocities,  Macharet et al. have studied the perimeter defense problem with multi-tasking defenders \cite{macharet2020adaptive}.  Here an adaptive partitioning scheme was used based on the estimated intruder's arrival distribution at the perimeter; also,  in that partition, one defender was assigned to capture multiple intruders. In \cite{bajaj2022competitive}, a single defender operating inside and outside the conical territory was used to protect a  conical territory from multiple radially incoming intruders with constant velocities. These preliminary works on multi-tasking defenders were limited to the case of specific (circular) territory and non-manoeuvring intruders.

\section{Solution Approaches for PDP}  \label{sec:PDP}
This paper proposes a defense strategy for the perimeter defense problem. First we briefly explain the PDP problem and then discuss the solution approach.  Before getting into the problem, we list a few notations and symbols used in the rest of this paper.  
 
\begin{enumerate}[\indent {}]
\item $ \Omega $ : the territory 
\item $\partial \Omega $ :  perimeter (boundary)of the territory 
\item $\Omega^P $ : priority region  
\item $\Omega^M $ :  monitoring region  
\item $D_i$ :  $i^{th}$ defender ;
\item $ I_j $ : $j^{th} $ intruder;
\item $ V $ : speed
\item $\gamma$ :  defender to intruder's speed ratio;
\item $\beta$ :   safety factor;
\item $\omega$ :  angular velocity;
\item $\bm{p}_i^D \in  \mathbb{R}^2$ :   position of defender $D_i$;
\item $\bm{p}_j^I \in  \mathbb{R}^2$ :   position of  the intruder $I_j$;
\item $\bm{p}_j^{\bot} \in \mathbb{R}^2$ :  projected point of the $I_j$ on the perimeter;
\item $\bm{p}^R_i \in  \mathbb{R}^2 $ :  location of the $i^{th}$ reserve station  ;
\item $\zeta $ :   priority factor;
\item $\zeta^M $ :  monitoring factor ;
\item $\bm{p}_j^T \in  \mathbb{R}^2$ :  arrival location of the intruder $I_j$ (i.e. task $T_j$);
\item $t_j $ :    time of arrival of the intruder $I_j$ ;
\item $T_j = T_j(\bm{p}_j^T, t_j)$ :  task defined by  the intruder $I_j$.
\end{enumerate}

\subsection{Problem formulation for perimeter defense problem  }
In a perimeter defense problem, while operating on and inside the territory, a team of defenders (with varying number of members) protects a  convex territory from intruders before they intrude the territory.
\begin{figure}[thb!]
    \centering
    \includegraphics[width=0.99\linewidth]{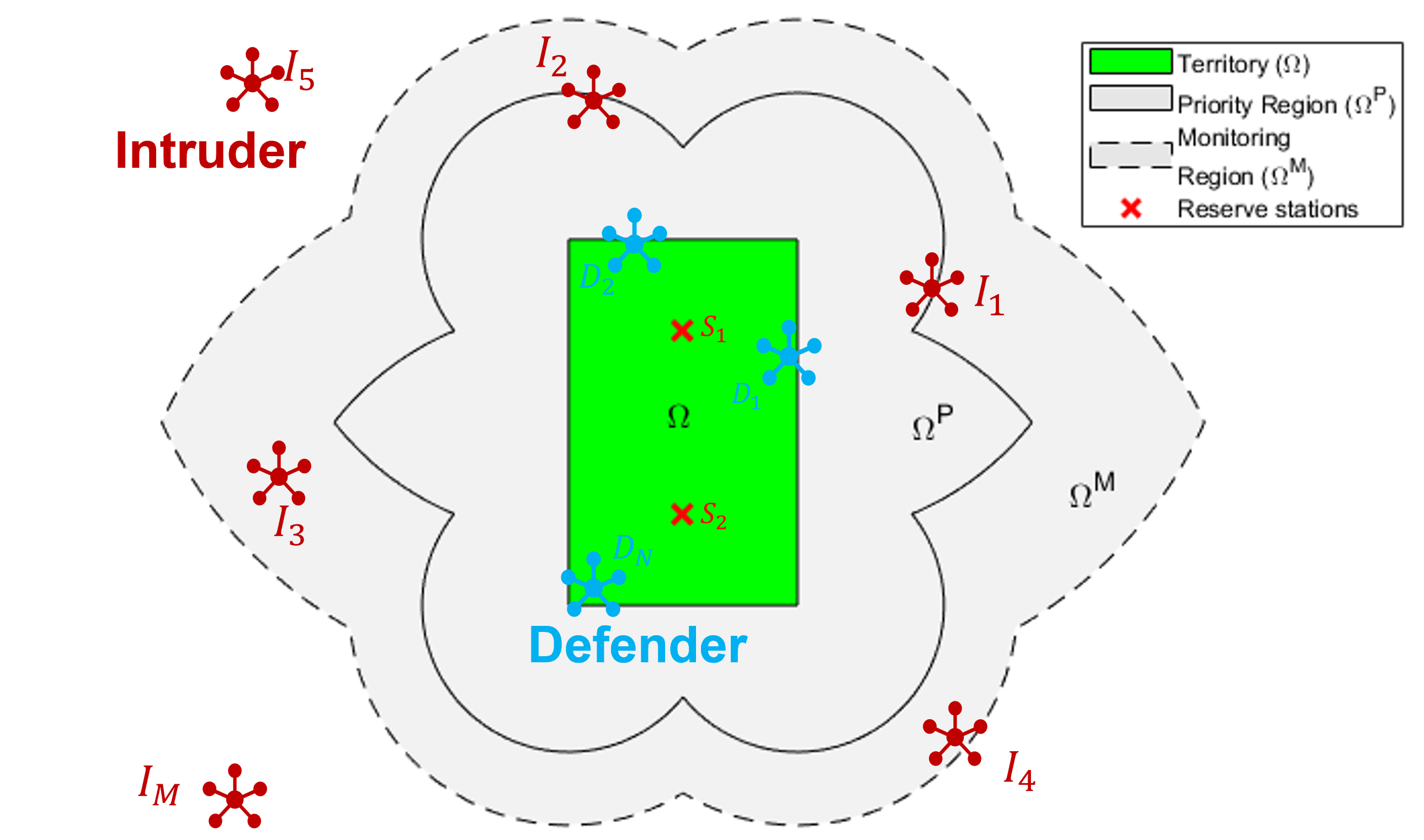}
    \vspace{5pt}
    \caption{Perimeter defense problem with reserve stations, priority and monitoring regions }
    \label{fig:Problem_formulation}
\end{figure}

A typical perimeter defense problem is shown in Fig. \ref{fig:Problem_formulation}. In the figure, the   safety-critical infrastructure exists inside a convex-shaped  region, called the territory and is denoted by $\Omega$, and $\Omega \in  \mathbb{R}^2 $. The perimeter of the territory is denoted by $\partial\Omega$. Fig. \ref{fig:Problem_formulation} also shows reserve stations,  priority region ($\Omega^P$), and monitoring region ($\Omega^M$) which are explained in next.  
Consider a team of defenders $ \{ D_1,D_2,\cdots,D_N \}$ protects the territory from the intruders denoted as $\{ I_1,I_2,\cdots,I_M \}$. 
Note that, the number of intruders is more than the number of defenders $\left(M>N\right)$ but defenders have a speed advantage $V_{max}^D>V_{max}^I$. The PDP problem formulated above is the same as given in \cite{velhal2022dynamic}. For brevity of space, details of the kinematic equations for the defenders and intruders are provided in the supplementary file.

Each defender is equipped with a sensor to detect the position and velocity of objects (intruders) in the region covered by its sensing radius $R_s$. The team of defenders collaboratively monitors the neighborhood around the territory to detect potential intruders. The approach developed here quantifies this neighbourhood region which needs to be continuously monitored and this region is referred to as the minimum monitoring region and denoted as  ($\Omega^M$); it is shown in Fig. \ref{fig:Problem_formulation} by grey shaded region inside the dotted boundary.

The defenders use a dynamic size of the team (varying numbers in which reserve defenders can be added or removed. The reserve stations are placed inside the territory as shown in Fig. \ref{fig:Problem_formulation}. It is assumed that the reserve stations have sufficient defenders available to be deployed based on the need, and these defenders can be deployed from reserve station whenever required. Defenders can neutralize the intruder by approaching closer to the intruder while the intruder is outside the territory. The neutralizing condition of an intruder $I_j$ by a defender $D_i$ is mathematically given as,
\begin{align} 
Q(I_j) = \left\{ D_i      \Big| \  {\|( \bm{p}_i^D -  \bm{p}_j^I ) \|}_2 \leq \epsilon        
   \  \& \  \bm{p}_j^I \in\Omega^{out} \right\}
\end{align}

Once a defender neutralizes an intruder it is free to neutralize another intruder. In this way, a defender can neutralize multiple intruders in a sequence. The observed intruders are predicted to intrude the perimeter based on their positions and velocities. The location where an intruder is predicted to intrude is referred as  the \textit{arrival location} and the predicted time of intrusion is referred as the \textit{arrival time}.

The objective of a perimeter defense system is to neutralize all the intruders before they enter the territory using an optimal number of defenders. The team size of defenders can vary as per the need to neutralize multiple intruders one after another given by the trajectory (sequence) computed by the task assignment algorithm.

\subsection{Dynamic Resource Allocation with Spatio-Temporal Multi-Task Assignment (DREAM) approach and its limitations}
The perimeter defense problem was addressed by formulating it as a spatio-temporal multi-task assignment problem \cite{velhal2021decentralized, velhal2022dynamic}. Each of the defenders in the team can observe the positions and velocities of the intruders, whenever the intruders enter into the sensing range of a defender.  Based on the position and velocity of the intruder $I_j$, the future trajectory of that intruder is predicted and used to compute the arrival point ($p_j^T$) and arrival time ($t_j$) of that intruder. Now, the intruder is predicted to reach the arrival point on the perimeter at its corresponding arrival time. Each intruder $I_j$, thus generates a spatio-temporal task for a defender to reach the arrival point of that intruder at the arrival time $T_j(p_j^T,t_j)$.  Next, a multi-task assignment algorithm was used to assign the defenders to these tasks. A detailed review on the multi-agent task assignment with temporal constraints is given in \cite{nunes2017taxonomy}

Each defender can be assigned to multiple spatio-temporal tasks and a defender will execute them in sequence (called as the trajectory).  In \cite{velhal2022dynamic}, the minimum number of defenders required to execute these multiple spatio-temporal tasks was computed. Also, a dynamic-sized team of defenders was used to improve the utilization of the defenders.

The formulation of PDP as a spatio-temporal multi-task assignment problem is based on the prediction of the arrival points and arrival times of the intruders. The reformulation of assignment problem at every time step with re-computation of the assignments helps to handle the dynamic actions of the intruders but still there is a possibility of intrusion.

\subsubsection{Point of arrival for highly manoeuvring intruder}
\begin{figure}[thb!]
    \centering
    \includegraphics[width=0.85\linewidth]{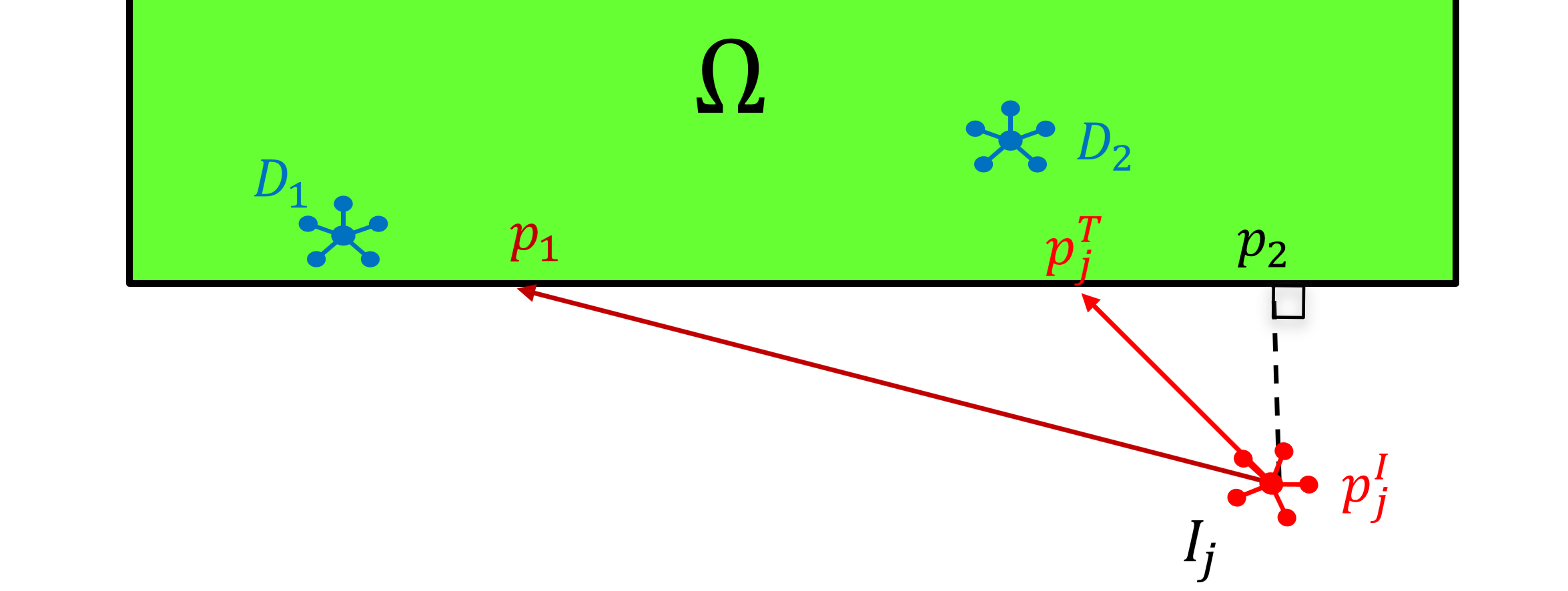}
    \vspace{1pt}
    \caption{ Typical scenario showing need of re-defining arrival location for highly manoeuvring intruder}
    \label{fig:agile_intru}
\end{figure}

The arrival point  ($p_j^T$) of an intruder $I_j$  is computed using predictions of its future trajectory based on its velocity and position. Now, for highly manoeuvring intruders predictions only based on velocity are often inaccurate. Fig. \ref{fig:agile_intru} shows one   such scenario where an intruder $I_j$ is at location $p_j^I$ and its velocity is directed towards, point $p_1$ on the territory. Defender $D_1$ is close to the point $p_1$ hence it will be assigned. The highly manoeuvring intruder can change its heading and it may move away from $D_1$. towards the projected point of $I_j$ on the territory  (i.e. $p_2$).  In such cases, the assignments will not be reliable, and they will change now and then after every heading change by the intruder.

This issue can be overcome by reducing the reliability on the velocity direction of the intruder.  The arrival location of the intruder is computed using its velocity direction and projection on the territory. The arrival location lies in between the projected point ($\bm{p_2}$) and extended velocity crossing point ($\bm{p_2}$).  The arrival location ($p_j^T$) of intruder is selected from the perimeter path $p_1,p_2$. $p_j^T$  is selected such that ratio of divided segments is equal to the ratio of distances  of intruder from $p_1$ and $p_2$.
Mathematically,
\begin{align} 
\frac{ {\| p_1,p_j^I \|}_2}{ {\|p_2,p_j^I \|}_2} = \frac{l(p_1,p_j^T )}{l(p_2,p_j^T )} 
\end{align}
where $l(a,b)$ denotes the distance along perimeter from point $a$ to point $b$.

\subsubsection{Issues in multi-task approach}
\begin{figure}[b!]
    \centering
    \includegraphics[width=0.85\linewidth]{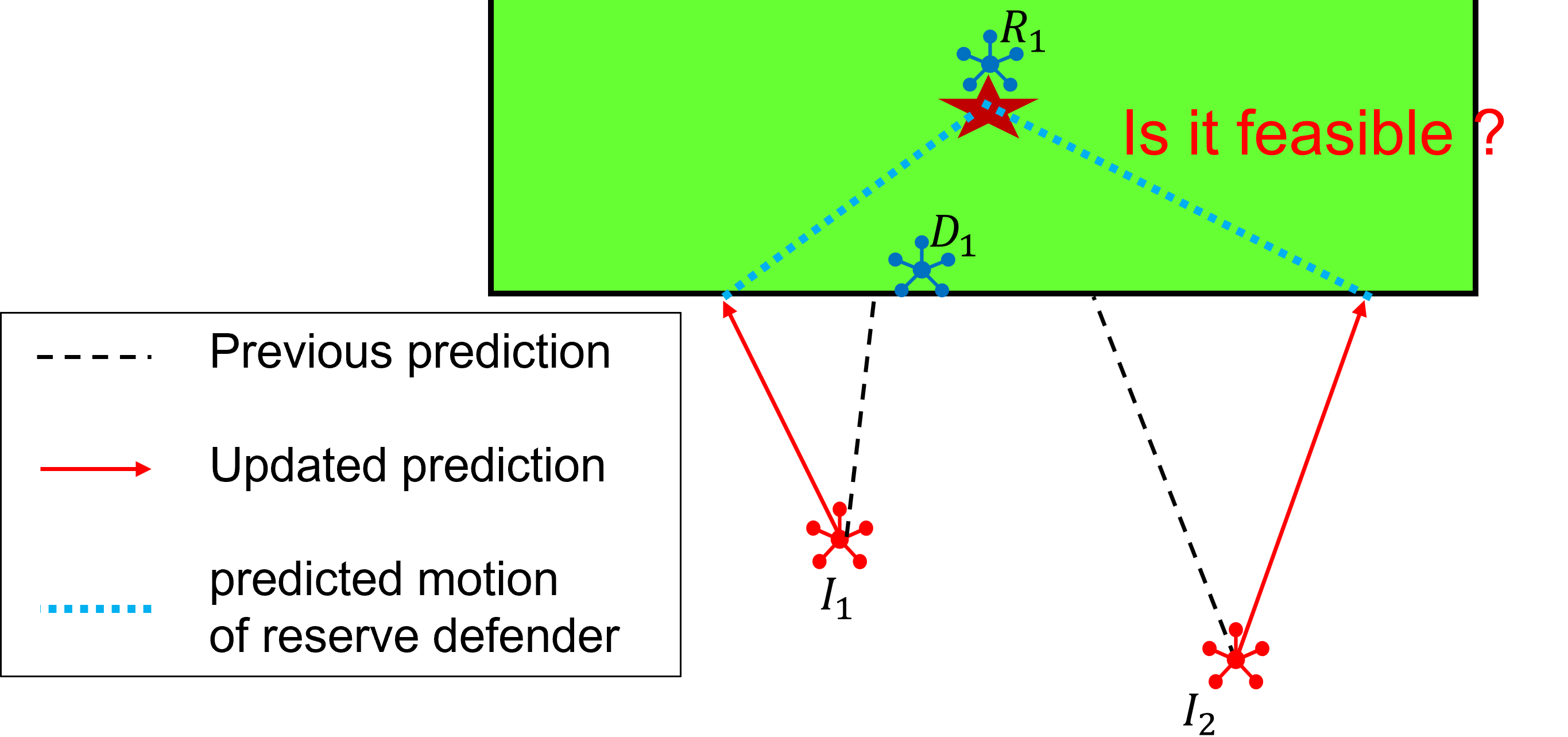}
    \vspace{1pt}
    \caption{ Issues in multi-task assignments for highly manoeuvring intruders  }
    \label{fig:failure_case1}
\end{figure}

The DREAM \cite{velhal2022dynamic} approach was proposed for the PDP problem to use an optimal number of defenders to neutralize all the intruders. This increases the success rate but still there are failures. The major reason for the failure of DREAM is in handling multiple intruders, who are close to each other by one defender. The Fig. \ref{fig:failure_case1} shows a typical scenario for the failure case.

Consider, defender $D_1$ is assigned to multiple intruders with a trajectory $\mu_i = \{I_1,I_2\}$,  based on their previous predicted trajectories. Now, in the next time step, the intruders have changed their heading angles. The changes in the heading angle of an intruder may lead to an infeasible problem as shown in Fig. \ref{fig:failure_case1}. In the previous step, both intruders are feasible for $D_1$ but they become infeasible now. A reserve defender needs to be called for neutralizing them.  An important question that arises here is: what is the guarantee that either or any of the intruders is feasible for the reserve defender $R_1$? If both intruders are very close to the perimeter, then the reserve defenders may not be able to neutralize either one of them and $D_1$ will neutralize only one intruder and one intruder will enter the territory. This paper addresses these questions with the idea of the priority of intruders defined using the priority region. The brief idea of P-DREAM is described next.

\subsection{Defense strategy in the P-DREAM approach}
It is important that a reserve defender should be called when an intruder is feasible to it, otherwise the reserve defender is helpless. This paper proposes a priority region around the territory such that every intruder in this priority region is given priority in task assignments and assigned as a first task.  Let us define a priority region ($\Omega^P$) around the territory such that if intruders enter this region, a defender must be assigned to neutralize that intruder on priority. The intruder inside the priority region is called as a prioritized intruder. The defender must neutralize the prioritized intruder as the first task.  The intruders outside the priority region can be assigned 
either as a first or a subsequent neutralization task.

All intruders in the priority region are assigned as first tasks. If an intruder changes its heading angle, the defender assigned to it can handle it because there will not be any other priority intruder assigned to that defender. If an intruder is outside the priority region and changes the heading direction, then it is not an immediate threat as that intruder is sufficiently away from the territory.

At any given time, the given team of defenders may or may not be sufficient to neutralize all the intruders. The reserve defenders need to be deployed whenever necessary. A new defender will be added to the team whenever neutralization of all intruders becomes infeasible with the given team of defenders. A new defender is added from the reserve stations. The use of prioritized multi-task assignment will guarantee that unassigned intruder is feasible for the reserve defender. A defender is removed from the team whenever it is unassigned and not required for the monitoring task.

Along with the intruder neutralization tasks, defenders need to monitor the neighbourhood of the territory to detect and track potential intruders. In general, the neighbourhood is a very vague concept and saying monitoring the neighbourhood does not really quantify the neighbourhood. The quantification of the monitoring region is an important design problem in PDP. Once the monitoring region is computed then one can find   the minimum  defenders required for the monitoring task.  Main aspects of designing the defense strategy for PDP are

\begin{enumerate}[a)]
\item Design the optimal locations for the reserve stations 
\item Computation of the priority region around the territory, with in which intruders should be given priority and assigned only as a first task.
\item Quantification of the neighbourhood which must be monitored, and determine the minimum number of defenders required for monitoring. 
\end{enumerate}
The next section presents the problem formulations to compute the design parameters listed above.

\section{Static Optimization Problem Formulations for Determining the Critical Parameters for PDP} \label{sec:static}
This section presents the formulations for the static parameter computations for a perimeter defense problem. For a given territory, speed ratio and the sensing range the following formulation provides a framework to compute the required number of reserve stations, their optimal locations (such that reserve defenders can reach the perimeter with  shortest  distance), priority region, minimum monitoring region and number of defenders required for monitoring.
\begin{algorithm}[h!]
 \caption{   Design of static layout   } \label{algo:design_steps}
\begin{algorithmic}[1]
\State {\bf{input} Convex territory, sensor range} 
\State {Select the number of reserve station } \label{algo:lable_initilaize}
\State {Compute the location for reserve stations } \label{algo:label_reserve_station}
\State {Compute the priority region}
\State {Compute the monitoring region}\label{algo:label_monotoring_region}
\State {Compute the minimum number of defenders and their sensing range required for monitoring } 
\If {computed sensing range and minimum number of defenders are infeasible}
\State increase the number of reserve station   
\State Go to step \ref{algo:lable_initilaize} \label{algo:label_iter}
\EndIf 
\State {Design is complete}
\end{algorithmic}
\end{algorithm}

The steps involved in this design problem are described in algorithm \ref{algo:design_steps}. For a given convex territory, the number of reserve stations are selected heuristically first. Then one needs to find the locations of these reserve stations. Based on the locations and number of reserve stations, one should then compute the priority and monitoring regions.  For the computed monitoring region, minimum number of defenders required for monitoring is then computed along with the required sensing range for monitoring. If the computed values for the sensing range and the number of defenders is feasible, then the design is complete. If those values turn out to be infeasible (outside the desired limits), then one needs to repeat the above steps again by increasing the number of reserve stations till one achieves a feasible solution.

\subsection{Determination of optimal locations for the reserve stations}
A reserve station should be placed inside the territory, from which a reserve defender can be deployed to handle the tasks when they become infeasible for the current team of defenders. The intruders can be neutralized when they come close to the perimeter and hence the reserve defenders need to travel to the arrival locations of the intruders. This arrival location can be any point on the perimeter. Hence the reserve stations should be placed such that the worst-case distance that needs to be travelled by the defenders to reach the perimeter is minimum. The number of reserve stations is a design parameter and is optimized in an iterative way to reduce the area of the monitoring region.  

\begin{figure}[htb!]
    \centering
    \begin{subfigure}{0.48\linewidth}
    \centering
        \includegraphics[width=0.95\linewidth]{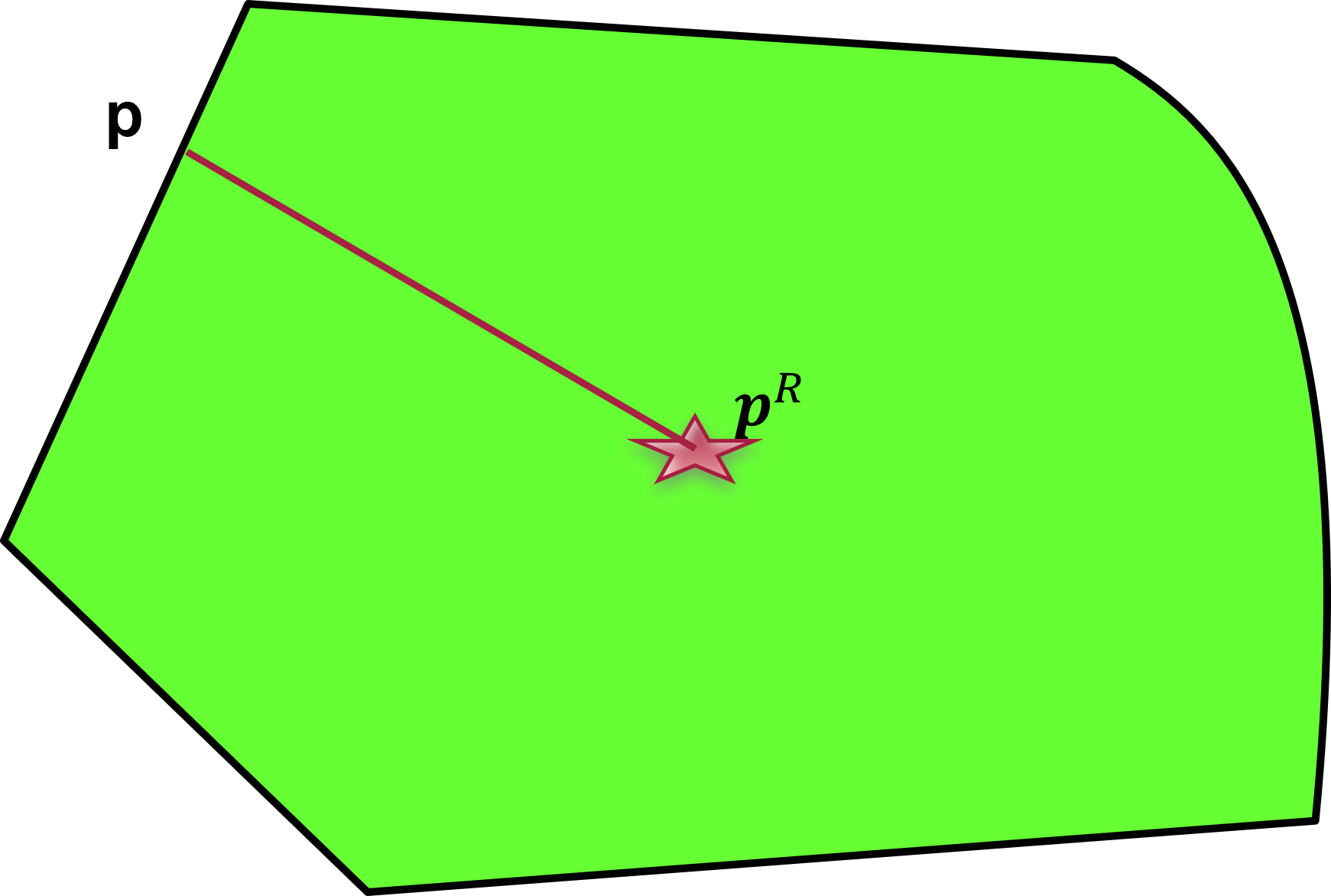}
        \subcaption{}
        \label{fig:Reserve_station_one}
    \end{subfigure}%
    \quad
    \begin{subfigure}{0.48\linewidth}
    \centering
        \includegraphics[width=0.95\linewidth]{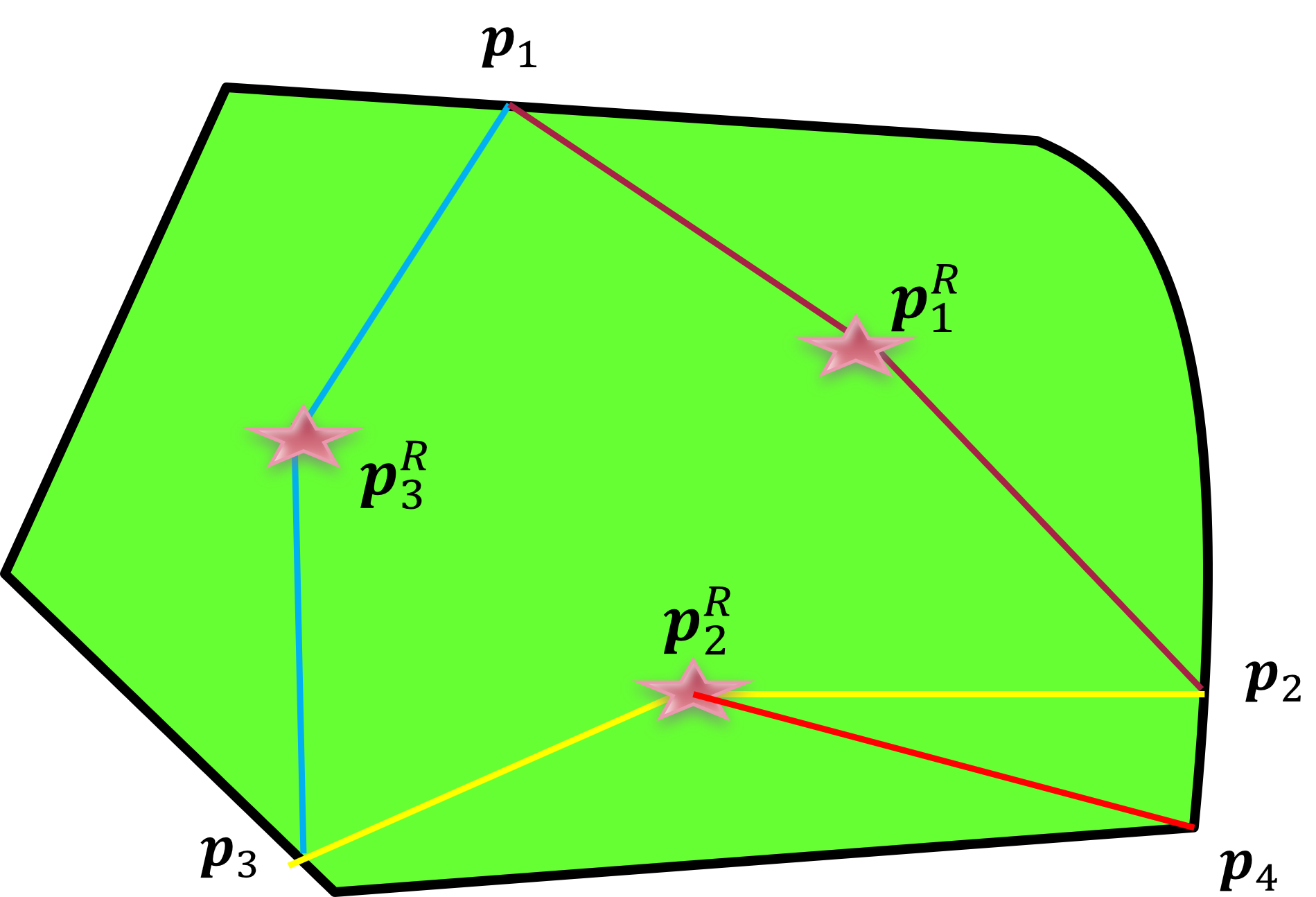}
        \subcaption{}
        \label{fig:Reserve_station_multi}
    \end{subfigure}
    \vspace{5pt}
    \caption{{Illustration of the optimal placement of reserve stations}
        \textnormal{ (a)single reserve station (b)multiple reserve stations }}
    \label{fig:Reserve_station}
\end{figure} 

Fig. \ref{fig:Reserve_station} shows  a  geometric interpretation of the above  formulation as a  mini-max problem. As a first step, the location for a single reserve station is computed and then it is extended   to the case of multiple reserve stations. For a single reserve station, as shown in Fig. \ref{fig:Reserve_station_one}, the reserve station is located at $\bm{p}^R$  such that, the maximum distance to any point $\bm{p}$ on the perimeter is minimized.  Mathematically this can be put as,  
\begin{align}
\min_{ \bm{p}^R \in \Omega  }    \max_{ \bm{p} \in \partial\Omega } {\|(\bm{p}  - \bm{p}^R ) \|}_2  \label{eq:minimax1}
\end{align}

For multiple reserve stations, the reserve stations are located such that the distance from the closest reserve station is minimized. Consider fig, \ref{fig:Reserve_station_multi},  where  three reserve stations are located such that, they divide the boundary in three parts. The portion of the boundary from points $\bm{p}_1$ to $\bm{p}_2$, in a clockwise direction is always closer to  the reserve station $\bm{p}_1^R$.  One needs to compute the locations of the reserve stations such that the  boundary is divided into sub-parts and the maximum distance from the reserve station to the corresponding  portion of the boundary is minimized. The reserve stations  $ \{ R_1,R_2,..,R_{N_R } \} $ are located such that the  maximum distance to every point on  the perimeter from the respective closest reserve station  is minimized. Mathematically this is expressed as 
\begin{align}
\argmin_{\bm{p}^R \in \{ \bm{p}^R_1, \bm{p}^R_2,.., \bm{p}^R_{N_R } \} }    \max_{ \bm{p} \in \partial\Omega } \left \{ \min_{i} {\|( \bm{p}  - \bm{p}^R_i ) \|}_2  \right \} \label{eq:minimax}
\end{align}

In eq \eqref{eq:minimax} the term in the curly bracket indicates that the reserve defender is selected from the closest reserve station. Now the maximization function considers all the points on the perimeter and then considers the worst-case scenario. The outer minimization function will operate over the positions of the reserve stations which minimizes the worst-case scenario.  

Next, the formulations for computing the priority and monitoring region are presented based on the above chosen locations of the reserve stations.

\subsection{Computation of the priority region }
The intruder close to the perimeter must be assigned as the first task. This priority is given to the intruder because if it is not assigned as the first task, it may result in failure due to the maneuverability of intruders assigned as the subsequent tasks. The priority region is defined around the territory such that any intruder outside the region is always feasible for whatever the manoeuvres undertaken by the intruder. If  an intruder is inside the priority region, to guarantee the neutralization of that intruder, that intruder is assigned on priority as a first task and a dedicated defender will be assigned to neutralize that prioritized intruder.
\begin{figure}[t!]
    \centering
    \includegraphics[width=0.99\linewidth]{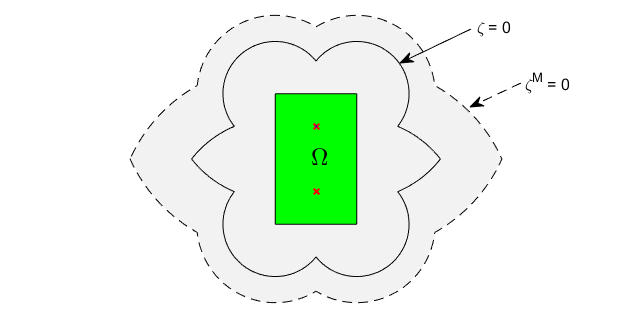}
    \caption{Illustration of priority and monitoring region}
    \label{fig:design_monitoring_region}
\end{figure}

The priority region is computed using the reserve stations in the territory. The region is computed such that the reserve defender should be able to neutralize the intruder. Let us consider a territory with $R_N$ reserve stations.
Suppose that the intruder enters at location $s$ at time $t_s$, if a reserve defender reaches the location $s$ before time $t_s$, then the defense is successful. 
Let us define the cost $J$ as the differential time to reach the location $s$; mathematically,
\begin{align} 
 J(s,  {\bm{p}}^I_j,V^D_{max},V^I_{max}  ) &=    \frac{ {\| {\bm{p}}^I_j -s \|}_2 }{ V^I_{max}} - \frac{ {\| {\bm{p}}^R_\star -s \|}_2 }{ V^D_{max}}
 \end{align}
where, ${\bm{p}}^R_\star$ is the closest reserve station from $s$.

The intruder chooses the location $s$ such that the intruder  can reach that location earlier than the defender. Let us define a priority factor $\zeta$ as the value which maximizes the scaled value of the $J$.
\begin{align}
\zeta({\bm{p}}^I_j,\gamma) &=  V^D_{max}  \max_{s \in \partial \Omega}{   J(s,  {\bm{p}}^I_j,V^D_{max},V^I_{max})}  \nonumber \\
\zeta({\bm{p}}^I_j,\gamma) &=  \max_{s \in \partial \Omega}{  \Big{\{}    \gamma  {\| {\bm{p}}^I_j -s \|}_2    - {\| {\bm{p}}^R_\star -s \|}_2   \Big{\}} }   
\end{align}
 
For an intruder with speed ratio of $\gamma$, at position ${\bm{p}}^I_j$ if $\zeta({\bm{p}}^I_j,\gamma) \le 0$ then the reserve defender can neutralize that intruder. Hence intruder is not prioritized. If $\zeta({\bm{p}}^I_j,\gamma) \ge 0$ then that intruder is infeasible for all reserve defenders. 

Now, the priority region is defined outside the territory as the  region where the priority factor is greater than zero.
\begin{align}
\Omega^P(\gamma) = \left\{ \bm{p}  \  \Big| \  \zeta(\bm{p},\gamma) \ge 0 \ \& \  \bm{p} \notin \Omega     \right\} \label{eq:Omega_P}
\end{align}

One can guarantee that every intruder outside the priority region is always feasible. Every intruder in the priority region should be neutralized on priority. To enforce this, the cost of the tasks are modified and is explained in section \ref{sec:dynamic}.

\subsection{Computation of the monitoring region }
Based on  the motivation of the priority region, the monitoring region  in the neighbourhood of territory will be  quantified. The monitoring region is defined as the minimum  region in the neighbourhood which has to be always monitored such that, if any intruder entering this region is detected and monitored, then that intruder can be neutralized with guaranteed success.
 
Fig. \ref{fig:design_monitoring_region} shows both the priority and monitoring regions. The monitoring ratio is defined   incorporating the uncertainty in the velocity measurements and the worst-case scenarios of the speed of intruder and also the sensing range.  A safety factor of  $\beta$, ( $\beta > 1$) is added in the computation of  the intruder's reaching time. This will allow tracking of intruder sufficiently earlier and well before it enters the priority region.

The differential time cost with safety is computed as
 \begin{align}
 J^M(s,\beta,{\bm{p}}^I_j,V^D_{max},V^I_{max}) = \frac{ { \| {\bm{p}}^I_j -s \|}_2 }{\beta V^I_{max}} - \frac{ {\| {\bm{p}}^R_\star -s \|}_2}{V^D_{max}}
\end{align}
 where, $\beta > 1$, ${\bm{p}}^R_\star$ is the closest reserve station from $s$. 
 
The monitoring factor ($\zeta^M$) is defined as that value which maximizes the scaled value of the $J^M$.
\begin{align}
\zeta^M({\bm{p}}^I_j,\gamma,\beta ) &=  V^D_{max}  \max_{s \in \partial \Omega}{   J^M(s,{\bm{p}}^I_j, V^D_{max}, V^I_{max}, \beta )} \nonumber \\
\zeta^M({\bm{p}}^I_j,\gamma,\beta ) &=  \max_{s \in \partial \Omega}{  \Big{\{}     \frac{\gamma }{\beta } {\| {\bm{p}}^I_j -s \|}_2    - {\| {\bm{p}}^R_\star -s \|}_2   \Big{\}} }    \label{eq:Omega_M}
\end{align}

Now the monitoring region ($\Omega^M$) outside the territory is defined as that area  where the monitoring factor is greater than zero.
\begin{align}
\Omega^M(\gamma,\beta) = \left\{ \bm{p}  \  \Big| \  \zeta^M(\bm{p},\gamma,\beta) \ge 0 \ \& \  \bm{p} \notin \Omega    \right\}
\end{align}
 using \ref{eq:Omega_M} and \ref{eq:Omega_P} 
 \begin{align}
\Omega^M(\gamma,\beta) = \Omega^P(\gamma / \beta )
\end{align}
 
Note that, the safety factor ($\beta$) is a design parameter and depends on the sensing range ($R_s$). The safety factor is chosen such that the furthermost point in the monitoring region can be sensed with the given sensor range while the defender is at a favourable location on the perimeter.

\subsection{Computation of minimum number of defenders required on the perimeter}
Once the monitoring region is computed, one can then compute the minimum number of defenders required for  monitoring . The solution   is similar to the approach used in the computation of the optimal locations for the reserve stations. In the computation of the optimal locations of reserve stations, the number of reserve stations was pre-decided;  but here the number of defenders required for monitoring is computed based on a given sensor range.

\begin{figure}[h!]
    \centering
    \includegraphics[width=0.95\linewidth]{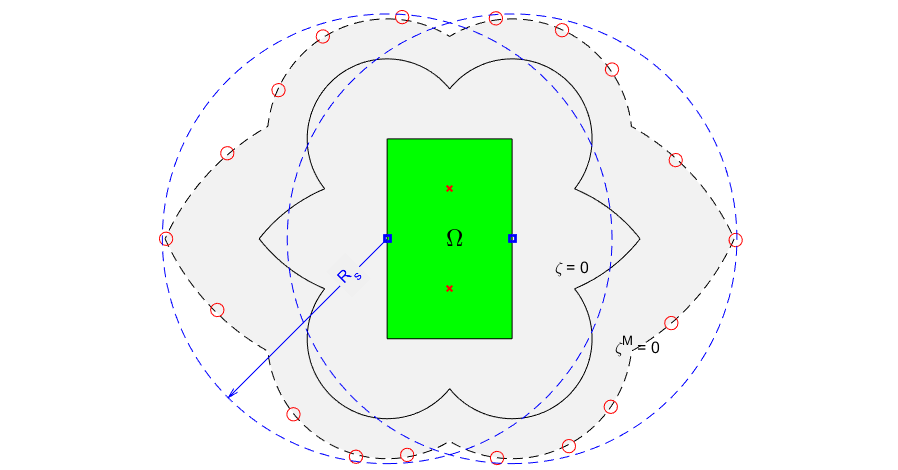}
    \vspace{0pt}
    \caption{ Illustration of minimum defenders required for monitoring of critical points.  \textnormal{ The sensor defenders are denoted by blue square and its sensing range is shown by dashed blue circle} }
    \label{fig:min_defender}
\end{figure}

Consider the monitoring region as the boundary and one needs to place the defenders on or inside the territory such that, collectively  they cover  the monitoring region. For this purpose and also for reducing the required sensing range, the distance from the monitoring boundary and the defenders should be minimized. Also, every point on the monitoring boundary should be monitored by the closest defender. Hence,  above the mini-max optimizing problem can be  defined as follows
\begin{subequations} 
  \addtocounter{equation}{-1}
  \begin{align}
  &\argmin_{\bm{p}^D \in \{ \bm{p}^D_1, \bm{p}^D_2,.., \bm{p}^D_{N } \} }    \max_{ \bm{p} \in \partial\Omega^M } \left \{ \min_{i} {\|( \bm{p}  - \bm{p}^D_i ) \|}_2  \right \} \label{eq:minimax_monitor} \\
& \qquad \qquad \text{such that,  } \bm{p}_i^D \in  \Omega \   \forall i 
\end{align}
\end{subequations} 

This problem computes the optimal locations for the given $N$ defenders while minimizing the maximum distance from the monitoring boundary (i.e. minimizing the sensing range required). 
The minimum sensing range required ($R_s^{min}$) is computed as 
\begin{align}
R_s^{min} =  \max_{ \bm{p} \in \partial\Omega^M } \min_{i} {\|( \bm{p}  - \bm{p}^D_i ) \|}_2 
\end{align}

\begin{algorithm}[htb!]
 \caption{ Minimum team of defenders required for monitoring } \label{algo:opti_sensor}
\begin{algorithmic}[1]
\State  {Initialize with  defenders N = 1 }
\State {Solve the optimization problem defined by eq. \eqref{eq:minimax_monitor} } \label{algo:label:IP1}
\State {Compute minimum sensing range ($R_s^{min}$) required }
\If { $R_s^{min} > R_s $ }
\State N = N + 1 
\State Go to step \ref{algo:label:IP1}
\EndIf 
\State {Solution is obtained, $N_{min} = N$}
\end{algorithmic}
\end{algorithm}

The solution to this problem may give an unrealistic sensing range. One needs to solve the problem by increasing the number of defenders until the solution results in a realistic sensing range. Also, if the territory is very large, onboard sensors on the defenders will not be able to monitor the monitoring region. One needs to either increase the sensor range or reduce the monitoring region.  The monitoring region can be reduced by increasing the number of reserve stations inside the territory. One needs to repeat the static design steps for an increased number of reserve stations as explained in step \ref{algo:label_iter} in algorithm \ref{algo:design_steps}. The whole procedure has to be done iteratively as all are closely interrelated. 

All the above optimization problems are static problems and can be computed offline for a given territory. These optimized parameters  are  used for the  operations of PDP. The formulation and working of guaranteed capture of multiple intruders in PDP is explained in the next section.

\section{Prioritized Multi-Task Assignment with Dynamic Resource Allocation} \label{sec:dynamic}

This section presents the mathematical formulation for P-DREAM  wherein the cost is modified to incorporate the prioritized intruder and the dynamic resource allocation algorithm is used to compute the optimal resources required for protecting the territory against highly manoeuvring intruders.

\subsection{Cost function}
The cost function is almost same as that was discussed in \cite{velhal2022dynamic}. The only change is made to enforce priority intruders as the first assignment by considering those that are infeasible in the subsequent task and correspondingly the subsequent task's cost is set to a large value $\kappa$. For completeness, the cost function is briefly described below. 

If the arrival location of an intruder can be reached by a defender on or before the arrival time of that intruder, then the intruder is feasible for that defender; otherwise, it is infeasible for that defender.  

The $C_{ij}^f$ is the cost for a defender $D_i$ to neutralize the intruder  $I_j$ as the first task. For a feasible intruder, the cost is proportional to the distance that needs to be traveled by a defender from its own position to the arrival location of the intruder. For an infeasible intruder, the cost is selected as a large value $\kappa$. Mathematically, the first cost is given as
\begin{align}
 &  C^f_{ij} = \begin{cases}  \alpha { \|( \bm{p}_j^T - \bm{p}_i^D )\|}_2  & \text{if } \dfrac{ { \|(  \bm{p}_j^T - \bm{p}_i^D )\|}_2  }{V_{max}^D}  \le  t_j  \\  
\kappa &  \text{ otherwise}
\end{cases} \label{eq:cost_DT}   \\  
&\qquad   \text{for } i \in \mathcal{I} = \{1,2,\cdots,N\} ,\quad   j \in \mathcal{J} = \{1,2,\cdots,M \}    \nonumber 
\end{align}
where $\alpha$ is a scaling factor proportional to the fuel consumed by the defender to travel a unit distance, and $\kappa$ is a large, fixed cost for an infeasible intruder.

The $C_{kj}^s$ is the cost for a defender to neutralize the subsequent intruder $I_j$ just after the neutralization of the intruder $I_k$. The defender will neutralize the intruder $I_k$ at location $\bm{p}_k^T$ at time $t_k$. after the neutralization of the intruder $I_j$, the defender should reach the arrival location of the intruder $I_j$ from $\bm{p}_k^T$ within the  available time $ t_j - t_k $. First we compute the minimum time required to travel this distance and then check for the feasibility of the task in a sequence. The minimum time required for a defender to reach the arrival location of the intruder $I_j$ from the location $\bm{p}_k^T$ is computed as
$$ t_{k,j}^{min} =   {\|( \bm{p}_j^T - \bm{p}_k^T  )\|}_2   / {V_{max}^D} $$

For a time-feasible intruder $I_j$, the cost $C_{kj}^s$  is the scaled distance between the location of the defender after neutralization of  $I_k$ and the arrival location of $I_j$. If the arrival time of $I_j$ is greater than that of $I_k$ and intruder $I_j$ is forward time-infeasible, then the cost is selected as $\kappa$. To guarantee the neutralization of the intruder, the intruders inside the priority region must be assigned as a first task. Hence the subsequent task for the intruder in priority region is considered as infeasible and the value set to $\kappa$.

If the arrival time of $I_j$ is lower than that of the neutralized intruder $I_k$, it is impossible to neutralize an intruder in the past from the future; hence, the cost is selected as $\infty$.
\begin{align}
 C^s_{kj} = \begin{cases}\alpha{\|( \bm{p}_j^T - \bm{p}_k^T)\|}_2 & \text{if } t_{  k,j}^{min} \le ( t_j - t_k  ) \\
\kappa & \text{if }    t_{  k,j}^{min} > ( t_j - t_k )> 0  \\
\kappa & \text{if }  I_j \in \Omega^P   \\
\infty & \text{if } t_j \le t_k 
\end{cases}  \label{eq:cost_TT}    \\
   \text{for } k \in \mathcal{K} = \{ 1,2,\cdots,M-1 \} \quad   j \in \mathcal{J} \qquad \nonumber 
\end{align}

\subsection{Optimization problem} 
A task assignment algorithm   assigns defenders to the neutralization tasks. A defender  will execute the tasks in a sequence and we denote the sequence assigned to the defender $D_i$ by $\mu_i$. Here,  the problem of computing a sequence $\mu_i$ has been converted to compute  each move of one defender from one location to another  and then  combining all the moves one can get the sequence of tasks. Each defender computes its sequence such that it executes all time-constrained neutralization tasks while minimizing the distances travelled. The first decision variable $\delta^f_{ij}$ is used to denote that either a defender moves from position $\bm{p}^D_i$ to the position $\bm{p}^T_j$ on or before time $t_j$. The subsequent decision variable   $\delta^s_{kj}$ is used to denote that either a defender moves from its previous task position $\bm{p}^T_k$  after time $t_k$ to the next task position   $\bm{p}^T_j$ on or before time $t_j$.

The integer programming problem is defined as
\begin{subequations} 
  \addtocounter{equation}{-1}
\begin{align} \label{eq:integer_prog}
     \min_{\delta^f_{ij} \ \delta^s_{kj}}  & \sum_{i\in \mathcal{I} }  \sum_{j  \in  \mathcal{J}    }  C^f_{ij} \delta^f_{ij}  +  \sum_{k\in \mathcal{K} }    \sum_{j  \in   \mathcal{J}    }  C^s_{kj} \delta^s_{kj}  \ \  \\ 
 {\rm s. \ t.} \   
 & \delta^f_{ij} \in \{0,1\}\qquad \forall (i,j) \in {  \mathcal{I} \times {\mathcal{J}  } } \label{eq:cost_cond_a} \\
 & \delta^s_{kj} \in \{0,1\}\qquad \forall (k,j) \in {  \mathcal{K}   \times {\mathcal{J} } } \label{eq:cost_cond_b} \\
 &\sum_{i \in \mathcal{I}}  \delta^f_{ij} + \sum_{k \in {\mathcal{K}  } }  \delta^s_{kj} = 1  \quad  \forall j \in  {\mathcal J} \label{eq:cost_cond_c} \\ 
 &\sum_{j \in {\mathcal{J} }}  \delta^f_{ij}   \le 1    \quad  \forall  i \in   \mathcal{I}     \label{eq:cost_cond_d}  \\
 &  \sum_{j \in {\mathcal{J} }}  \delta^s_{kj} \le 1   \quad  \forall   k \in  {\mathcal{K}}       \label{eq:cost_cond_e}  
\end{align}  
\end{subequations}

All tasks must be assigned to exactly one defender (from one location); this constraint is given by \eqref{eq:cost_cond_c}. A defender can move to at most one task location just after  the  current task and this is constrained by eq \eqref{eq:cost_cond_d} and  \eqref{eq:cost_cond_e}.

\subsection{ P-DREAM approach  }
The optimal number of defenders required for the execution of spatio-temporal tasks has been computed using the DREAM \cite{velhal2022dynamic} algorithm. The same algorithm is modified (given in algorithm \ref{algo:opti_resource}) to handle prioritized intruders. The algorithm consists  of two parts, first is to compute the required minimum number of defenders to execute the  spatio-temporal tasks and assign them to those spatio-temporal tasks. The second part is to remove   an unassigned defender from the team when it is not required for monitoring.

\begin{algorithm}[t!]
\caption{ Optimal resource allocations algorithm } \label{algo:opti_resource}
 \begin{algorithmic}[1]
\State  {Initialize with  given $N$ defenders and $M$ intruders and formulate the tasks with time constraints}
\State For all critical points, find the defenders monitoring them
\State {Compute the cost for tasks } \label{algo:label:TOCM}
\State {Solve the optimization problem (eq. \ref{eq:integer_prog}) } \label{algo:label:start_addition}
\If {assignments have cost equal to $\kappa$ }
\State  $q$ = number of assignments with cost $\kappa$
\State Find reserve stations close to infeasible task's locations 
\State Add $q$ defenders from selected reserve stations;
\State $N = N+q $
\State  Go to step   \ref{algo:label:TOCM}
\EndIf  \label{algo:label:end_addition}
\While { All critical points are not monitored }
\State {Add a defender to team from reserve station close  to \phantom  a   \phantom . that unmonitored critical  point. $N = N +1 $ }
\EndWhile
\If {All defender are  assigned to task}
\State  Solution is obtained . Go to step \ref{algo:label:stop}
\Else
\For  {all unassigned defenders}
\State Find the set ($\mathcal{P}^c$) $\mathcal{P}^c$ of critical points only   \phantom a  \phantom  . \phantom  .  \phantom  .  \phantom . \phantom . \phantom . \phantom . monitored by  the unassigned defender.
\If {set $\mathcal{P}^c$ is empty }
\State Remove unassigned defender; $ N = N-1 $
\Else
\State  Defender is required for monitoring
\EndIf
\EndFor
\EndIf
\State {Stop} \label{algo:label:stop}
\end{algorithmic} 
\end{algorithm}  

The solution of the task assignment problem \eqref{eq:integer_prog} helps in computing the required number of reserve defenders. From the solution of \eqref{eq:integer_prog} one can compute the infeasible assignments as all of them will have cost equal to $\kappa$. The reserve defenders equal to the number of infeasible assignments are added to the team as described in steps \ref{algo:label:start_addition}-\ref{algo:label:end_addition}. Next, a defender will be added for the monitoring task if required for monitoring.

The unassigned defender is removed if not required for monitoring. Here, the monitoring region is defined quantitatively, so one can decide the critical points (consisting of kink points) for the monitoring region. Checking at those critical points, one can find whether an unassigned defender is required for monitoring or not?  For removal of unassigned defender $D_i$, first  all the critical points monitored by the defender $D_i$ are computed   given by the set $\mathcal{P}^c$. Now for all  the pointsin  the set  $\mathcal{P}^c$, one can check whether any other defender can sense that point or not? If all the points in $\mathcal{P}^c$ are at least monitored by a defender other than $D_i$, then an unassigned defender $D_i$ is not required for the  monitoring task and should be removed.

 \section{Performance Evaluation of P-DREAM} \label{sec:result}

The section presents the simulation results for the proposed P-DREAM approach for PDP.
To illustrate the P-DREAM approach, a convex polygon with vertices at $(20,0)$, $(10,30)$, $(-20,20)$, $(-30,-10)$, and $(0,-20)$ is considered as the territory to be protected. The area of territory is $1600 sq.m.$. Intruders are randomly trying to enter the territory, and their maximum velocity is $ V_I^{max} = 3m/sec$. Defenders have a maximum speed of $3m/sec$. The defender's speed is varied for studying the effects of speed ratio. All the simulations were conducted in MATLAB R2022a.

\begin{figure}[htb!]
    \centering
    \includegraphics[width=0.95\linewidth]{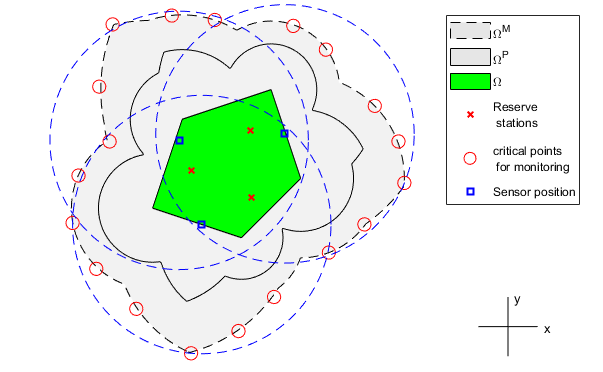}
    \caption{Illustration of PDP for computed static parameters with priority and monitoring regions \textnormal{(parameters $n_R=3$, $\gamma = 1$, $\beta = 1.33$, $N_{min} = 3$, and $R_s = 43.613m$ )} }
    \label{fig:critical_monitor_points}
\end{figure}
The static parameters for the given territory are computed as described in \ref{sec:static}. Fig. \ref{fig:critical_monitor_points} shows the territory with the designed static parameters. Fig. \ref{fig:critical_monitor_points} shows the priority region ($\Omega^P$) for the  territory with three reserve stations (and a speed ratio of one) in grey shaded region in a solid boundary. The monitoring region($\Omega^M$) for speed ratio, $\gamma = 1$, and safety factor, $\beta = 1.33$ is shown by a grey-shaded region in a dashed boundary. The critical points are marked by red circles. The minimum number of defenders($N_{min}$) required for monitoring is three. Fig. \ref{fig:critical_monitor_points} shows three defenders by blue squares and their sensing regions by blue circles. The detailed calculations for the above static design are provided in the supplementary materials.

\subsection{Working of prioritized multi-task assignment approach}
The first phase (static design) of the PDP design  computes  $N_{RS}=3$, $N_{min}=3$. Also, the priority and monitoring regions have been computed. Next, the second phase i.e. the working of the prioritized multi-task assignment approach is illustrated here.

\begin{figure*}[t!]
    \centering
\begin{subfigure}[thb!]{0.24\linewidth} 
    \includegraphics[width=1\linewidth]{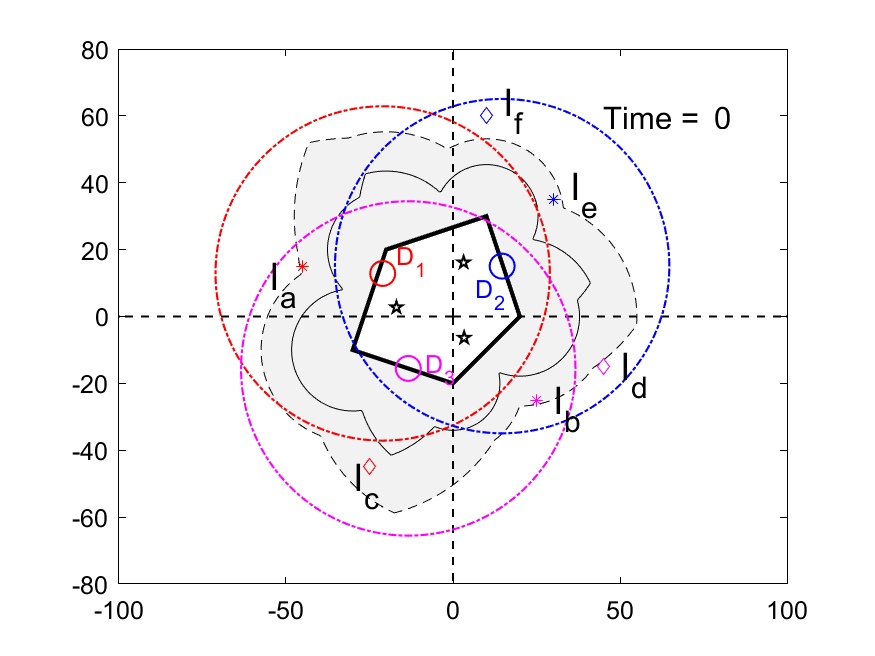}
    \caption{time = 0 sec  }
    \label{fig:sub1}
    \end{subfigure}
\begin{subfigure}[thb!]{0.24\linewidth}
    \includegraphics[width=1\linewidth]{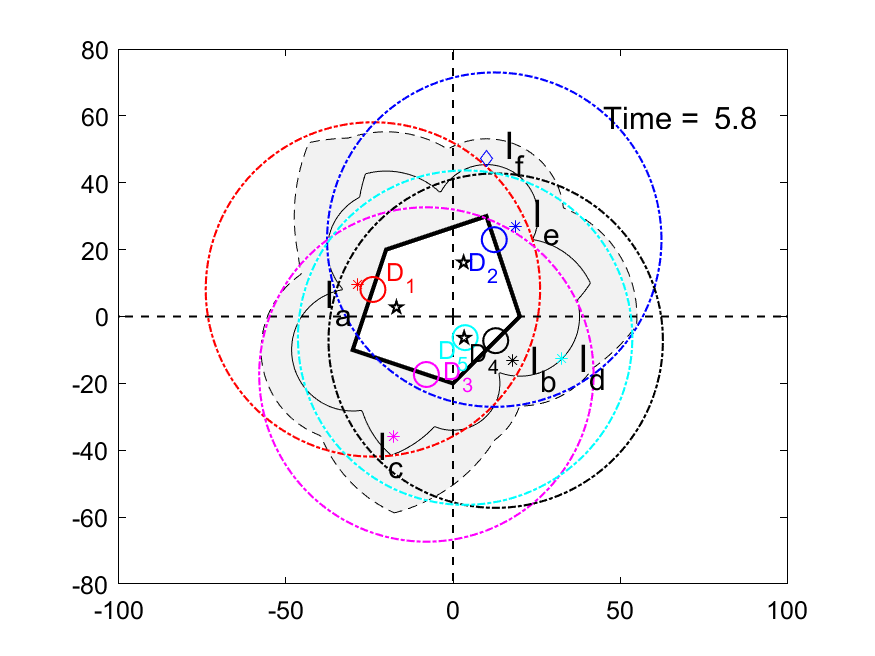}
    \caption{time = 5.8 sec  }
    \label{fig:sub2}
\end{subfigure} 
\begin{subfigure}[thb!]{0.24\linewidth}
    \includegraphics[width=1\linewidth]{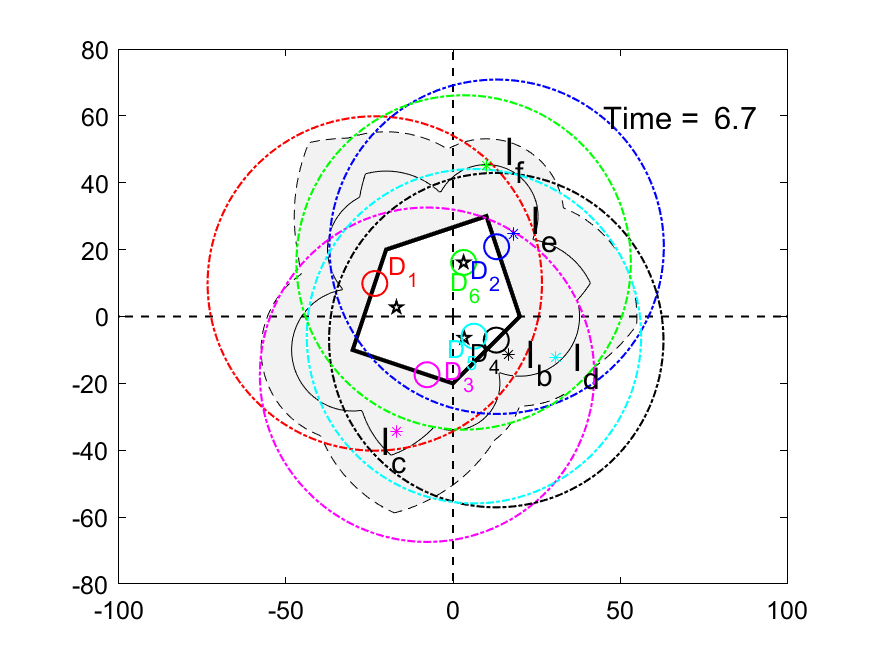}
    \caption{  time = 6.7 sec  }
    \label{fig:sub3}
\end{subfigure}  
\begin{subfigure}[thb!]{0.24\linewidth} 
    \includegraphics[width=1\linewidth]{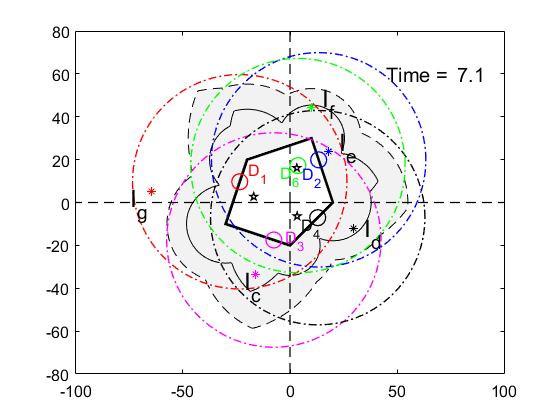}
    \caption{ time = 7.1 sec  }
    \label{fig:sub4}
    \end{subfigure}\\
\begin{subfigure}[thb!]{0.24\linewidth} 
    \includegraphics[width=1\linewidth]{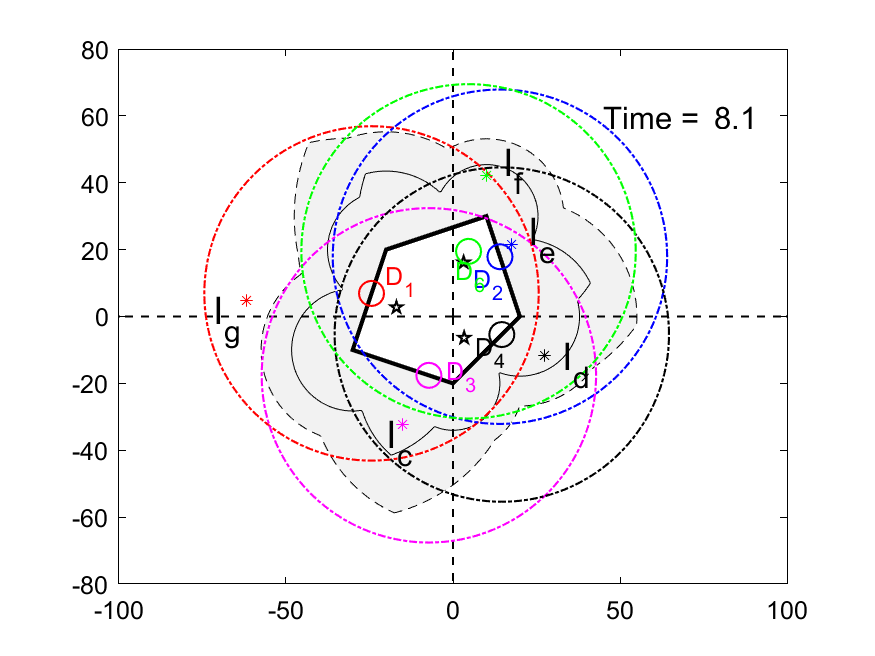}
    \caption{ t = 8.1 sec   }
    \label{fig:sub5}
    \end{subfigure} 
\begin{subfigure}[thb!]{0.24\linewidth}
    \includegraphics[width=1\linewidth]{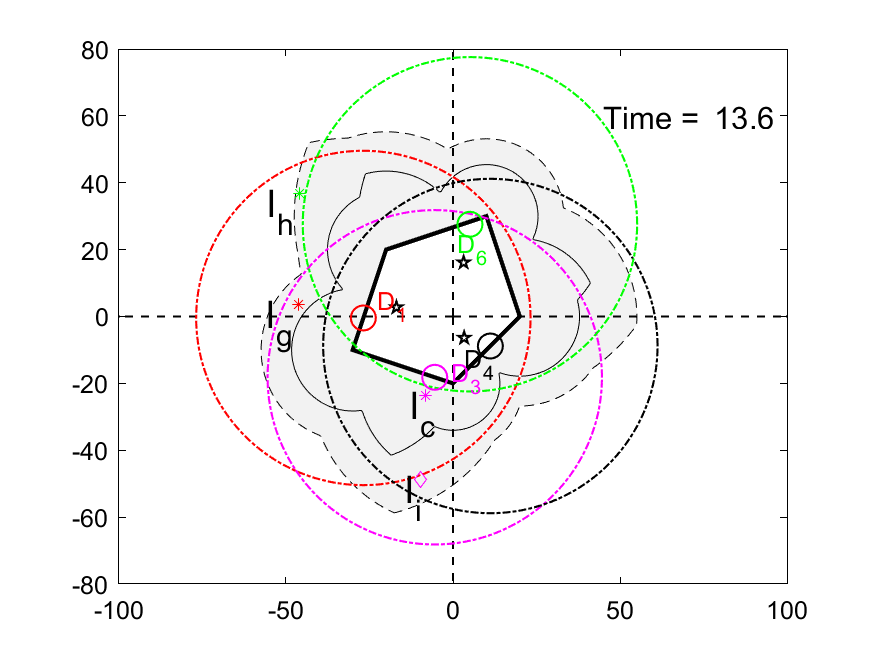}
    \caption{ time = 11.3 sec}
    \label{fig:sub6}
\end{subfigure} 
\begin{subfigure}[thb!]{0.24\linewidth}
    \includegraphics[width=1\linewidth]{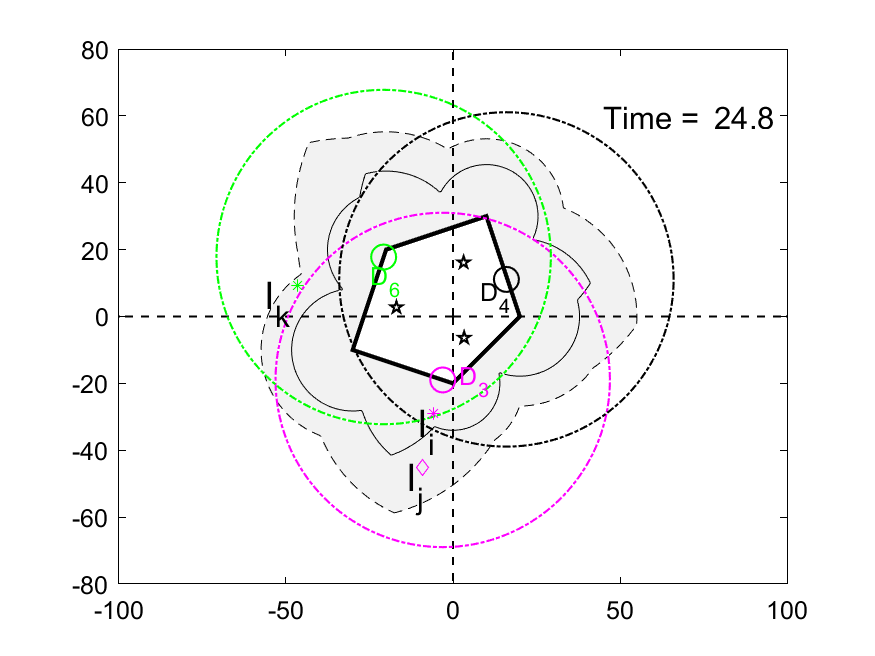}
    \caption{ time = 24.8 sec }
    \label{fig:sub7}
\end{subfigure}
\begin{subfigure}[thb!]{0.24\linewidth}
    \includegraphics[width=1\linewidth]{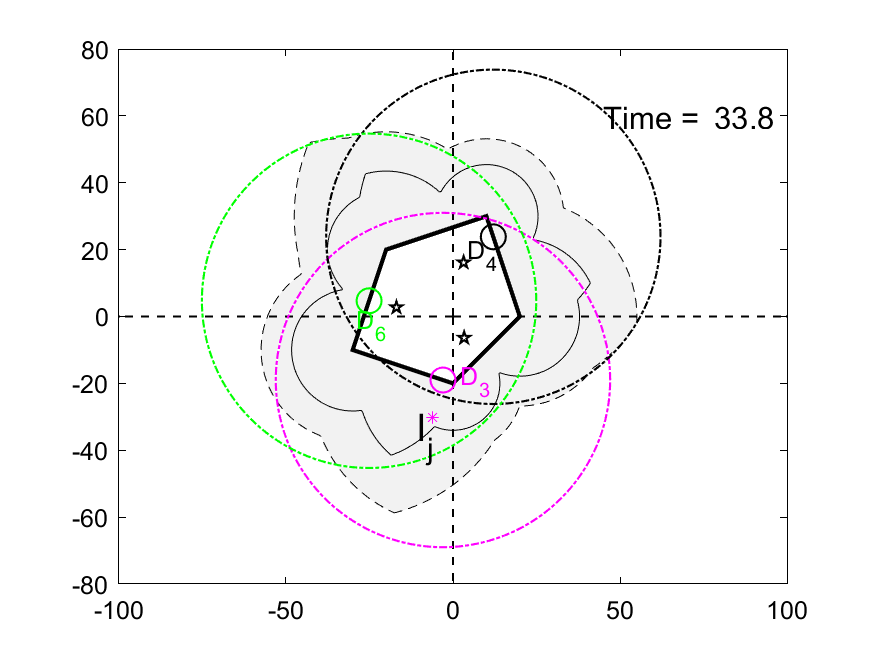}
    \caption{ time = 32.6 sec }
    \label{fig:sub8}
\end{subfigure}   
\caption{Snapshots of scenarios in PDP with P-DREAM}
\label{fig:Working_PDREAM}
\end{figure*}

Fig. \ref{fig:Working_PDREAM} shows snapshots of typical scenarios in PDP with P-DREAM simulations.   The convex territory is shown by bold black coloured solid lines. The three reserve stations inside the territory are marked with star symbols.  The priority region ($\Omega^P$) is denoted by a grey region inside the black solid lines and the monitoring region ($\Omega^M$) is denoted by the grey region inside dashed black lines. The initial condition is shown in Fig. \ref{fig:sub1}, with three defenders marked in different colours.  The assigned intruders are marked with the same colour as the defenders. The sequence of assigned intruders is shown by symbols *, $\diamond$, $\square$ for first, second and third respectively. A video to show the operation of defenders for neutralizing the 10 intruders for entire duration of 37.5 seconds was generated. The same is given in the supplementary material. The critical time instances from the attached video are discussed here to explain the working of the P-DREAM approach.

At t= 5.8 sec, as shown in Fig. \ref{fig:sub2}, $I_d$ enters the priority region, initially $I_d$ was assigned as a second task for $D_4$, after neutralization of $I_b$. But, now $I_d$ has entered the $\Omega^P$, it needs to be assigned as the first task. The available 4 defenders are not sufficient, hence a new reserve defender $D_5$ is added for neutralization of $I_d$ as the first task. Next, at t = 6.7 sec, as shown in Fig. \ref{fig:sub3}, $I_f$ has entered the priority region $\Omega^P$. A new defender $D_6$ is added to the team for neutralization of $I_f$ as the first task.  

At time = 7.0 sec, the $D_4$ captures $I_b$. Next, at time = 7.1 sec, as shown in Fig. \ref{fig:sub4}, the $D_4$ is reassigned to intruder $I_d$, and $D_5$ which was added specifically to neutralize the prioritized $I_d$ is unassigned and eventually removed without any neutralization task. But, addition of $D_5$ was necessary to guarantee the success. The intruders $I_b$ and $I_d$ operates sub-optimally and hence only $D_4$ is able to neutralize both. At time = 8.1 sec, as shown in Fig. \ref{fig:sub5}, the $D_2$ captures the $I_e$. Here one can see that $I_e$ was successful in deviating the $D_2$ away from the arrival point of $I_f$. At this instant, $I_f$ becomes infeasible for $D_2$ and any reserve defender from $RS_1$. But,  $D_6$ which was already added in team is able to tackle the $I_f$.

At time = 13.6 sec, $D_4$ neutralizes $I_d$. Here, $D_4$ is required for the monitoring, hence $D_4$ is kept in the team for monitoring as shown in Fig. \ref{fig:sub6}.  $D_6$ neutralizes $I_f$ at time = 12.5 sec. From the video, one can also observe that in during a time span of 12.5sec to 24.8 sec $D_6$ travels toward the neutralization point of $I_h$ and due to this some of the monitoring points are not monitored. Here the monitoring defender $D_4$ operates such that it maximizes the monitored region. Fig. \ref{fig:sub7}, shows the snapshot at time = 24.8 sec. One can observe that $D_4$ has moved upwards to increase the coverage. Also, after the neutralization of $I_h$ at time = 24.8 sec, $D_6$ and $D_4$ operate cooperatively to increase the coverage of the monitoring region. At time = 33.8 sec, as shown in Fig. \ref{fig:sub8}, defenders are able to monitor the entire monitoring region.  

P-DREAM approach adds and removes   defenders to the team whenever needed  to guarantee the neutralization of all intruders. One can observe that the proposed P-DREAM approach guarantees higher success rates but may require more resources (defenders). The trade-off between the guaranteed neutralization of all intruders and the resource utilization is studied using Monte-Carlo simulations and are presented next.

\subsection{Effectiveness of mission success of P-DREAM over DREAM approach}

For slowly manoeuvring intruders, a dynamically-sized team of defenders works efficiently compared to a fixed-sized team of defenders. This has been shown in DREAM \cite{velhal2022dynamic} but DREAM fails to handle highly manoeuvring intruders. The proposed P-DREAM approach here overcomes the issues of high manoeuvrability of intruders, and the same is shown below using a Monte-Carlo study.

\begin{figure}[tbh!]
    \centering
    \includegraphics[width=0.95\linewidth]{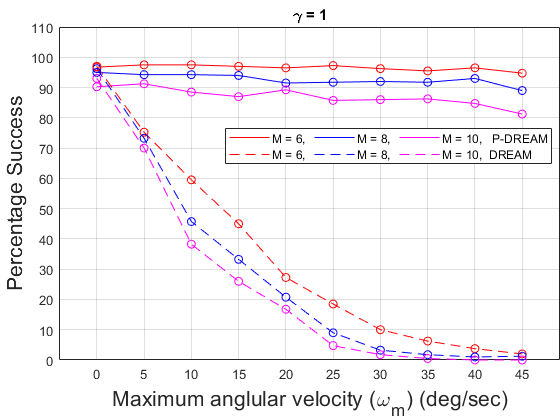}
    \vspace{1pt}
    \caption{Success percentage of P-DREAM and DREAM for different manoeuvring capacity of intruders   }
    \label{fig:Success_comparison}
\end{figure}

The P-DREAM and DREAM approaches are compared for the same territory with $3$ reserve stations and $3$ defenders on the perimeter for monitoring. Also, the monitoring task was removed (by selecting a large sensing range) and the intruder neutralization performance by the defenders is analyzed. Monte-Carlo simulations has been conducted for randomly incoming $15$ intruders with their respective manoeuvring rates. The system was initialized with three defenders and cases of $M = 6,8$, and $10$ intruders respectively, and a new intruder was added after the neutralization of an earlier intruder. A defense is successful only if all the intruders are captured before they enter the territory.  
The average success rates for total 400 simulation runs for both DREAM and P-DREAM are is presented in Fig. \ref{fig:Success_comparison}. 

One can observe that for manoeuvrability limits in the range of $0-45 \ deg/sec$,    the success percentage of P-DREAM is consistent and does not degrade with an increase in the manoeuvrability limits of intruders; whereas the success percentage of DREAM reduces and practically reaches zero for $\omega_m = 45 deg/sec$ whereas for P-DREAM it remains at $80\%$, $90\%$ and $95\%$ for $M = 10,8,6$ respectively.

In P-DREAM, the intruder's arrival location is computed based on the position and velocity of the intruder and the increase in manoeuvrability rates does not increase the prediction errors. Also, whenever an intruder enters the priority region, it is assigned as the first task. Hence, the P-DREAM approach provides consistent success rates.

\section{Conclusions}  \label{sec:conclusion}
A Priority-based Dynamic REsource Allocation with distributed Multi-task assignment (P-DREAM) approach for capturing highly manoeuvring intruders in a perimeter defense problem has been presented. The P-DREAM approach consists of two phases. In the first phase,  a  static optimization problem for a given convex territory to compute/design the parameters such as the number of reserve stations needed, their locations, priority regions, monitoring regions, and the minimum number of defenders required for monitoring is computed. These optimized parameters are then used in the framework/setup of the perimeter defense problem. 
In the second phase, for handling highly manoeuvring intruders, in the P-DREAM approach, a dedicated defender is assigned to each prioritized intruder by enforcing the prioritized intruder as a first task. The performance of P-DREAM has been illustrated in a simulated environment. The results show that the defender team defends the territory with consistent success against highly manoeuvring intruders. For the case of 3 active defenders with three reserve stations and a speed ratio of one against six active intruders with the manoeuvrability of $45deg/sec$, the success rate is in the range of $90-95\%$.

\bibliographystyle{IEEEtran}
\bibliography{main_bib.bib}

\vskip 0pt plus -1fil 
\begin{IEEEbiography}[{\includegraphics[width=1in,height=1.25in,clip,keepaspectratio]{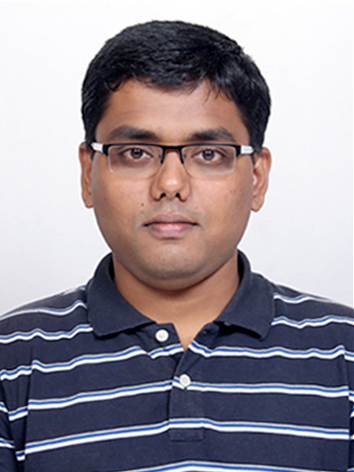}}]{Shridhar Velhal } is a Ph.D. student
in department of aerospace engineering at Indian Institute of Science (IISc), Bengaluru, India. 

His area of interest are in dynamics and control of autonomous systems, with a particular focus on multi-agent systems, territory protection games, spatio-temporal tasks, multi-task assignments and resource allocation.
\end{IEEEbiography}

\vskip -20pt plus -1fil
\begin{IEEEbiography}[{\includegraphics[width=1in,height=1.25in,clip,keepaspectratio]{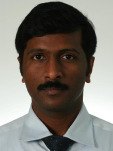}}]{Suresh Sundaram} (Senior Member, IEEE) received the Ph.D. degree in aerospace engineering from the Indian Institute of Science, Bengaluru, India, in 2005.

He is currently an Associate Professor with the department of Aerospace Engineering, Indian Institute of Science. From 2010 to 2018, he was an Associate Professor with the School of Computer Science and Engineering, Nanyang Technological University, Singapore. His research interests include flight control, unmanned aerial vehicle design, machine learning, optimization, and computer vision.
\end{IEEEbiography}

\vskip -20pt plus -1fil
\begin{IEEEbiography}[{\includegraphics[width=1in,height=1.25in,clip,keepaspectratio]{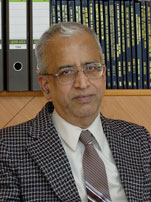}}]{Narasimhan Sundararajan} 
(Life Fellow, IEEE) received the Ph.D. degree in electrical engineering from the University of Illinois at Urbana-Champaign, Urbana, IL, USA, in 1971. 

From 1971 to 1991, he was with the Vikram Sarabhai Space Centre, Indian Space Research Organization, Trivandrum, India. Since 1991, he has been a Professor (Retd.) with the School of Electrical and Electronic Engineering, Nanyang Technological University (NTU), Singapore. Currently, he is a research staff in WIRIN project at IISc Bengaluru, India. His current research interests include spiking neural networks, neuro-fuzzy systems, and optimization with swarm intelligence. 
\end{IEEEbiography}

\newpage

\section*{ Supplementary materials} 
\subsection{Kinematic equations of motion for defenders and intruders }

The kinematic equations governing the motion of  defenders and intruders are described here.

\subsubsection{Defender}
Consider the $i^{th}$ defenders $D_i$, at  position   $\bm{p}_i^D=\left(x_i^D,y_i^D\right)$, with the velocity $V_i^D$, and the heading angle $\psi_i^D$. Its kinematic equations are as follows.
\begin{flalign}
&\dot{x}_{i}^{D} = V_{i}^{D} \cos\left(\psi_{i}^{D}\right),   \qquad  \dot{y}_{i}^{D} = V_{i}^{D} \sin \left(\psi_{i}^{D}\right),  \    \\[2pt]
& \text{ where, } \psi_{i}^{D} \in[0,2 \pi),  \ \    V_{i}^{D} \in\left[0, V_{max}^{D}\right],  \ \   \bm{p}_{i}^{D} \left( t_0 \right) \in \Omega   \nonumber  
\end{flalign}

\subsubsection{Intruder}
Consider  the $j^{th}$ intruder $I_j$ at  position  $\bm{p}_j^I=\left(x_j^I,y_j^I\right)$, with the velocity be $ V_j^I$ and the heading angle be   $\psi_j^I$ . The kinematic equations of $I_j$  are as follows:
\begin{align}
&\dot{x}_{j}^{I}=V_{j}^{I} \cos \left(\psi_{j}^{I}\right) \qquad  \dot{y}_{j}^{I}=V_{j}^{I} \sin \left(\psi_{j}^{I}\right)   \\[2pt]
& \text{where, } \psi_{j}^{I} \in[0,2 \pi), \ \ V_{j}^{I} \in\left[V_{\min}^{I_j}, V_{\max}^{I_j} \right], \   \bm{p}_{j}^{I}(t_0) \in \Omega^{out}    \nonumber   
\end{align}

\section*{ Static parameter computation}
The PDP design approach presented in this paper broadly consists of two phases, viz. the first phase is to come up with the design parameters for a given territory under the PDP design framework, and the second phase is to design the P-DREAM approach. 
To illustrate the P-DREAM approach, a convex polygon with vertices at $(20,0)$, $(10,30)$, $(-20,20)$, $(-30,-10)$, and $(0,-20)$ is considered as territory to be protected. The area of territory is $1600 sq.m.$. Intruders are randomly trying to enter the territory, and their maximum velocity is $ V_I^{max} = 3m/sec$. Defenders have a maximum speed of $3m/sec$. The defender's speed is varied for studying the effects of speed ratio.

For starting the design, as a first step  one needs to decide on the number of reserve stations required for the given territory. The guideline for selecting this number is to ensure it effectively reduces the size of the monitoring region (and also the priority region). Also, one should remember that any additional reserve station requires additional capital cost and this should be kept minimum. To achieve this, first, one has to compute the monitoring region by varying the number of reserve stations as these two are closely interconnected and the same is shown in the following subsection. 
 
\subsection{Computation of priority region, monitoring region, and number of reserve stations}

 \begin{figure*}[htb!]
    \centering
    \begin{subfigure}[thb!]{0.32\linewidth} 
    \includegraphics[width=0.98\linewidth]{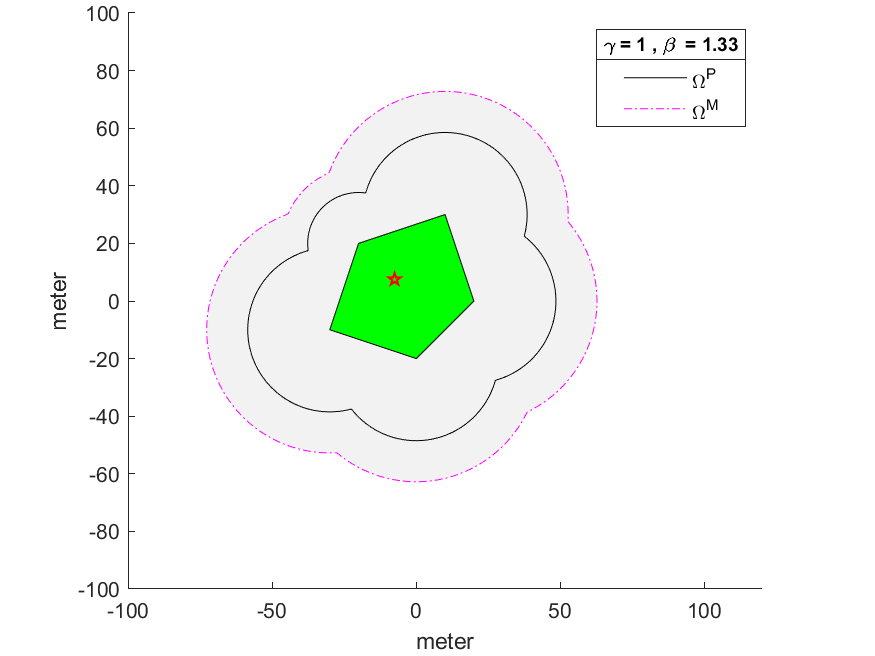}
    \vspace{0pt}
    \caption{$N_{RS} = 1$ \\ \ }
    \label{fig:priority_ploygon_1RS}
    \end{subfigure}
\begin{subfigure}[thb!]{0.32\linewidth}
    \includegraphics[width=0.98\linewidth]{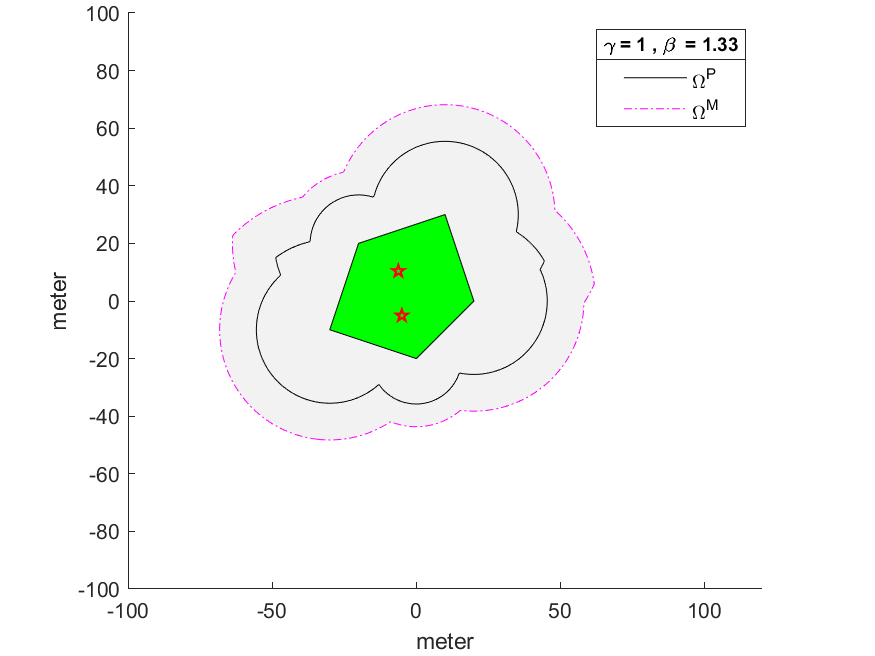}
    \vspace{0pt}
    \caption{  $N_{RS} = 2$ \\ \ }
    \label{fig:priority_ploygon_2RS}
\end{subfigure}
\begin{subfigure}[thb!]{0.32\linewidth}
    \includegraphics[width=0.98\linewidth]{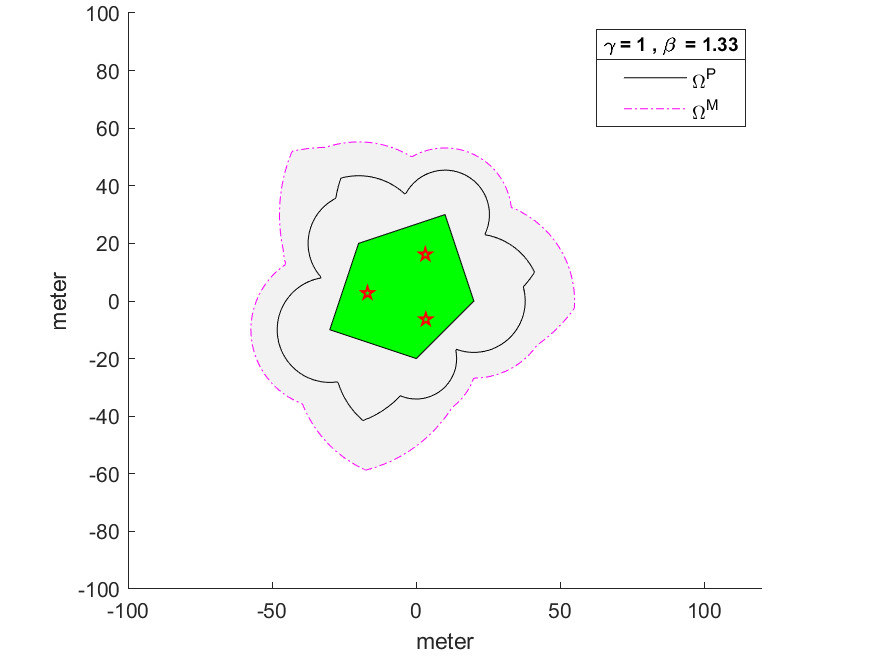}
    \vspace{0pt}
    \caption{$N_{RS} = 3$ \\ \ }
    \label{fig:priority_ploygon_3RS}
\end{subfigure}
\vspace{0pt}
\caption{The priority and monitoring regions for different reserve stations }
\label{fig:priority_ploygon}
\end{figure*}

The Fig. \ref{fig:priority_ploygon_1RS},\subref{fig:priority_ploygon_2RS},\subref{fig:priority_ploygon_3RS} shows the priority regions  ($\gamma = 1$) and monitoring regions ($\beta = 1.33$) for one, two, and three reserve stations respectively. . Both axes represent  the dimensions of the territory in meters.  From Fig. \ref{fig:priority_ploygon}, one can observe that the priority and monitoring region shrink as the number of reserve stations increases.

In Fig. \ref{fig:priority_ploygon}, the reserve stations are marked by a red cross (\textcolor{red}{$\times$}). 
For $n_{R} = 1$, the optimal location is computed by solving eq ~\eqref{eq:minimax1} and reserve station is located at $ ( -7.50,7.50)$. For $n_{R} = 2,3$ the optimal locations are computed by solving eq ~\eqref{eq:minimax}. This mini-max problem has many local solutions and hence the problem is solved multiple (1000) times with randomized initial conditions to obtain a global solution. Note that, Computation of optimal location for reserve stations is a static optimization problem and can be solved offline. The optimal locations obtained for $n_{R} = 2$ are $(-6.19,10.37) , (-4.99, -5.05)$. For $n_{R} = 3$, the reserve stations are located at $\bm{p}_1^R = (3.17,16.16) $, $\bm{p}_2^R = (3.33,-6.37)$ and $\bm{p}_3^R = (-16.92,2.74)$.

\subsubsection{Effect of speed ratio on the priority region}
  \begin{figure*}[htb!]
    \centering
    \begin{subfigure}[thb!]{0.32\linewidth} 
    \includegraphics[width=0.98\linewidth]{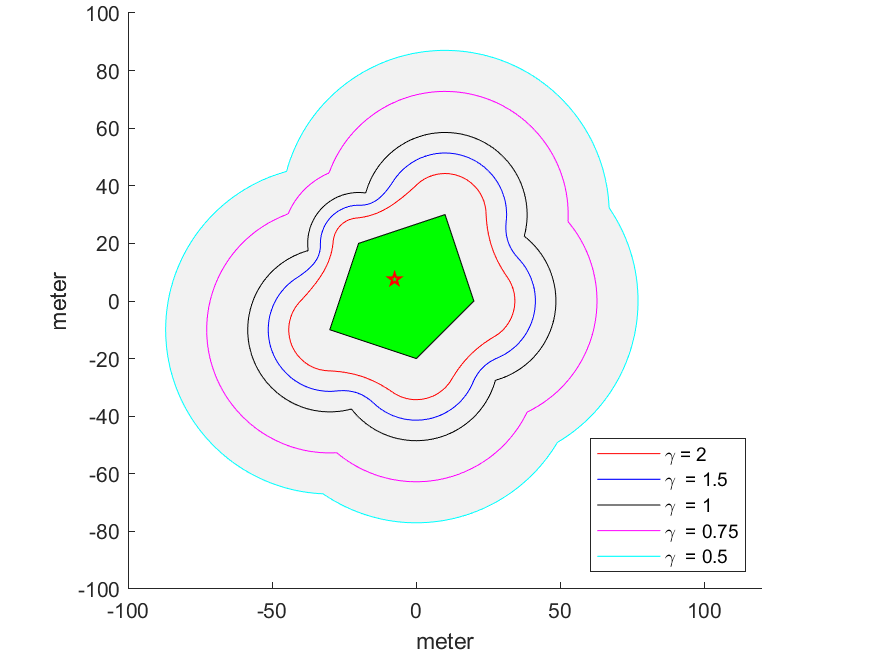}
    \vspace{1pt}
    \caption{Priority region for $N_{RS} = 1$ \\ \ }
    \label{fig:priority_ploygon_1RS_variations}
    \end{subfigure}
\begin{subfigure}[thb!]{0.32\linewidth}
    \includegraphics[width=0.98\linewidth]{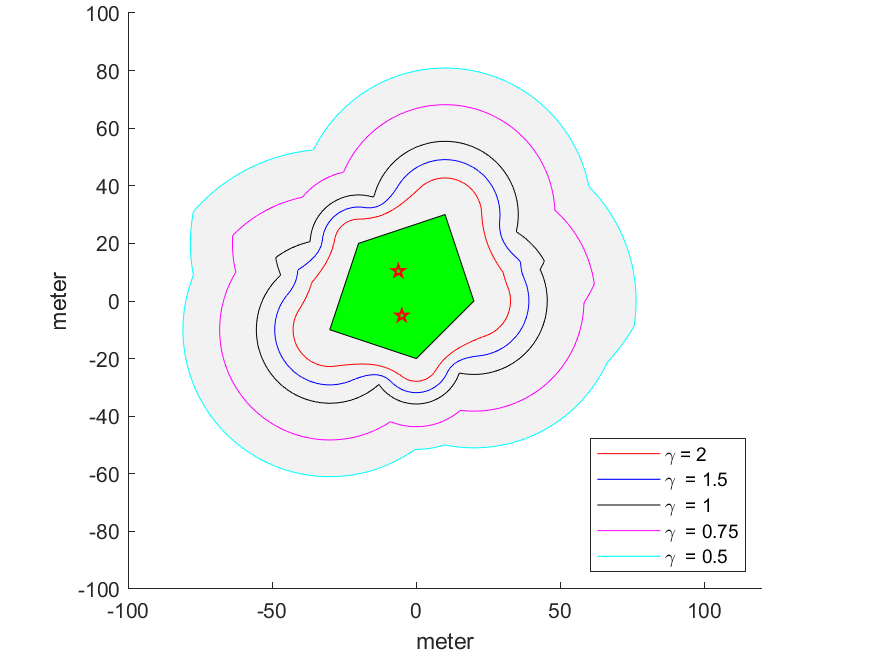}
    \vspace{1pt}
    \caption{ Priority region for $N_{RS} = 2$ \\ \ }
    \label{fig:priority_ploygon_2RS_variations}
\end{subfigure}
\begin{subfigure}[thb!]{0.32\linewidth}
    \includegraphics[width=0.98\linewidth]{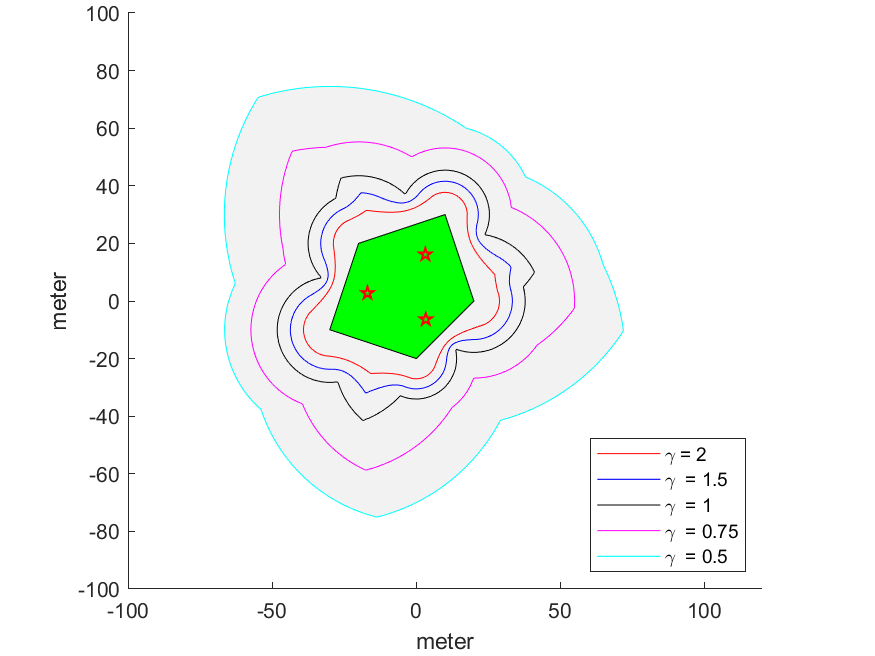}
    \vspace{1pt}
    \caption{Priority region for $N_{RS} = 3$ \\ \ }
    \label{fig:priority_ploygon_3RS_variations}
\end{subfigure}
\caption{The priority regions for different reserve stations and speed ratio}
\label{fig:priority_ploygon_variations}
\end{figure*}

The defender to intruder speed ratio is a critical parameter and affects the priority region. 

The Fig.\ref{fig:priority_ploygon_1RS_variations}, \subref{fig:priority_ploygon_2RS_variations},\subref{fig:priority_ploygon_1RS_variations} shows the priority region with one, two, and three reserve stations respectively.
Both axes are the dimensions of the territory in meters. In Fig.  \ref{fig:priority_ploygon_variations}, the reserve stations are marked by a red cross (\textcolor{red}{$\times$}). Each sub-figure shows the priority regions for the various defender to intruder speed ratio ($\gamma$).
From Fig. \ref{fig:priority_ploygon_variations}, one can observe that the priority region shrinks as the number of reserve stations increases. Also, as the speed ratio ($\gamma$) reduces the priority area increases.

The speed ratio is dependent on the max speed of the intruder. Speed ratio can be partially selected /designed by changing the maximum speed of the defenders. Here, we are considering a speed ratio equal to one. 
 \begin{figure}[t!]
    \centering
    \includegraphics[width=0.9\linewidth]{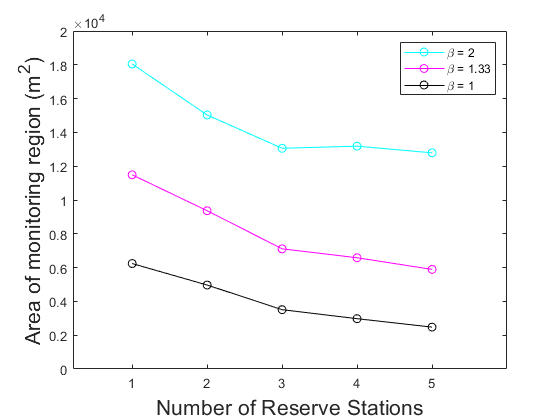}
    \vspace{0pt}
    \caption{Changes in monitoring area for different reserve stations}
    \label{fig:nrs_vs_Area}
\end{figure}
Fig. \ref{fig:nrs_vs_Area} shows the monitoring area required for  different number of reserve stations (for the cases where reserve stations are placed optimally).  The monitoring area reduces with an increase in the number of reserve stations for all safety ratios. Also, for a given number of reserve stations, as the safety ratio increases, the monitoring area also  increases. From Fig. \ref{fig:nrs_vs_Area}, one can see that the monitoring area reduces faster from one reserve station to three reserve stations. But after three reserve stations, the monitoring area  reduction rate is smaller  for any extra added reserve station. Hence,   the rest of the paper considers  three reserve stations case. 

\subsection{Computation of defender team size for continuous monitoring}

\begin{figure}[htb!]
    \centering
    \includegraphics[width=0.9\linewidth]{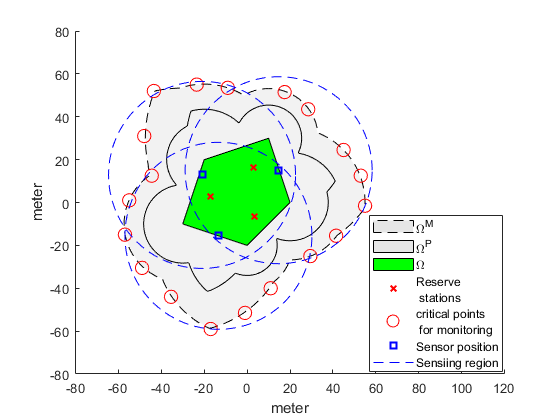}
    \vspace{0pt}
    \caption{minimum team of defenders required for monitoring (case: $n_R=3$, $\zeta^M = 0.75$, $N_{min} = 3$, and $R_s = 43.613m$ )  }
    \label{fig:critical_monitor_points_1}
\end{figure}
Consider a scenario with three reserve defenders,  speed ratio is 1, and the safety factor  $\beta = 1.33$ . 
Fig. \ref{fig:critical_monitor_points_1} shows the priority region for a territory with three reserve stations in grey shaded region in solid boundary. The monitoring region is shown by grey shaded region in dashed boundary. The one can identify the critical points needs to be monitored. The minimum number of defenders required for monitoring is computed by solving eq ~\eqref{eq:minimax_monitor}. Fig. \ref{fig:critical_monitor_points_1} shows three defenders with sensors and their sensing region by blue circles.

\begin{table}[htb!]
     \centering
      \setlength{\arrayrulewidth}{1.2pt}
      \caption{The sensing radius required for minimum number of defenders}
    \label{table:sensing_radius}
\begin{adjustbox}{max width=0.5\textwidth} 
\renewcommand{\arraystretch}{1.25}
\begin{tabular}{|l|r|r|r|r|r|}
\hline
        $N_{min}$          & 1 \quad \  & 2  \quad \    & 3  \quad \    & 4 \quad \    &5 \quad \  \\ \hline
        $R_s \ (m)$   & 61.3719 & 54.1359 & 43.6130 & 42.3745 & 42.3745  \\  \hline
\end{tabular}
\end{adjustbox}
  \vspace{0pt}
 \end{table}

The table \ref{table:sensing_radius} gives the sensing radius required for the different number of defenders to monitor the minimum monitoring area. The table shows that as the minimum number of defenders in a team increases the required sensor range reduces. But increasing the minimum number of defenders will help up to a certain limit, afterwards, there is no significant effect of the additional defender in the team. 
\textit{Depending on the available sensor radius, one needs to fix the minimum number of defenders required for monitoring.} Here we consider that the sensors have a range of $50m$. So,  the minimum number of defenders required in the team is three ($N_{min} = 3$).

\end{document}